\documentclass[aps,prl,twocolumn,amsmath,amssymb,floatfix,longbibliography,superscriptaddress]{revtex4-2}

\usepackage{color}
\usepackage{mathrsfs}
\usepackage{graphicx}
\usepackage{dcolumn}
\usepackage{bm}
\usepackage[breaklinks=true, pdftitle={}]{hyperref}
\usepackage[normalem]{ulem}
\usepackage{amsmath,amsfonts,amssymb,ulem}
\usepackage{epstopdf}
\usepackage{xcolor}
\usepackage[separate-uncertainty]{siunitx}

\newcommand{\black}[1]{{\color{black}#1}}

\usepackage{graphicx}
\begin{document}

\title{Radiation-induced electron spin polarization in ultrarelativistic kinetic turbulence}

\author{Peng Liu}
\affiliation{Institute of Theoretical Physics, Chinese Academy of Sciences, Beijing 100190, China}

\author{Karen Z. Hatsagortsyan}
\affiliation{Max-Planck-Institut f\"{u}r Kernphysik, Saupfercheckweg 1, 69117 Heidelberg, Germany}

\author{Christoph H. Keitel}
\affiliation{Max-Planck-Institut f\"{u}r Kernphysik, Saupfercheckweg 1, 69117 Heidelberg, Germany}
\author{Zheng Gong}
\email[]{zgong92@itp.ac.cn}
\affiliation{Institute of Theoretical Physics, Chinese Academy of Sciences, Beijing 100190, China}

\date{\today}
\begin{abstract}
Electron spin polarization in ultrarelativistic radiative plasma kinetic turbulence is investigated using particle-in-cell (PIC) simulations under highly magnetized conditions. We find that significant spin polarization can be sustained when the leptons undergo energetic photon emission accompanied by spin flips during the nonequilibrium turbulent evolution. 
\black{The correlation of spin signals and magnetic spectra with turbulent features allows us to access the fine structure of the charge-starved electromagnetic (EM) regime of kinetic turbulence, as well as to find a neat characteristic for the transition from density-dominated turbulence to the EM-regime. By analyzing the competitive processes of spin accumulation and the spin depolarization, we have identified two hitherto unexplored sub-regimes, dissipative and nondissipative, in the EM framework. We highlight a surprising finding
that in the EM regime the magnetic power-law spectrum develops a positive slope, in contrast to the commonly observed decaying spectrum.}
These results indicate the versatility of a spin-resolved study of plasma turbulence in extreme environments, such as magnetar inner magnetospheres.
\end{abstract}
\maketitle

Plasma kinetic turbulence (PKT) is a chaotic, multi-scale motion of a plasma where the physics can no longer be described by fluid dynamics. Here, the energy injected at macroscopic scales cascades down to kinetic scales through nonlinear wave-particle interactions~\cite{gary1993theory,horton1999drift,howes2008kinetic,schekochihin2009astrophysical,servidio2012local,grovselj2019kinetic,arzamasskiy2023kinetic,park2025kinetic}, leading to dissipation and nonthermal particle acceleration~\cite{Blasi2013,matsumoto2015stochastic,zhdankin2017kinetic,comisso2018particle,Zhdankin2019,ComissoAPJL2022,Vega2022,zhou2023electron,ewart2025relaxation,zhu2025polarization,gorbunov2025leaking}. 
Nonrelativistic PKT governs anomalous transport in magnetic confinement fusion~\cite{Jenko2000,Mattoo2012ETG,ongena2016magnetic}, drives Fermi-like stochastic acceleration in laser-produced plumes~\cite{Fiuza2020,Yuan2024,Liu2024WeibelTurb}, and regulates solar-wind heating in the heliosphere~\cite{Roberto2016solar,kawazura2020ion,Bowen2025solar}. 
In relativistic conditions, such as outflows from active galactic nuclei~\cite{Marscher2008AGN,Madejski2016,Bourne2017} or gamma-ray bursts~\cite{gamma_ray_burst,Meszaros2006,Burgess2020}, PKT mediates the conversion of ordered bulk energy into chaotic cosmic-ray motion, shaping the observed nonthermal spectra~\cite{Aab2020Observatory,Abdul2023Observatory}.

\black{In highly magnetized environments, such as pulsars~\cite{michel1982theory,uzdensky2014physical,philippov2022pulsar} and magnetars~\cite{kaspi2017magnetars,Parfrey2013MS,Li2016MS,Schoeffler2019MS}, particularly within the inner magnetosphere at tens of stellar radii above the magnetar surface, magnetic fields can reach $\sim 10^{10}$~G ~\cite{Thompson1995MNRAS,Yang2015Polarization}.
The magnetospheric perturbations are commonly triggered by crustal failures or starquakes~\cite{blaes1989neutron_starvation,LiMagnetar2016,YuanMagnetarQuake2020,Bransgrove2020QuakeAlfven},
and these catastrophic events launch large-amplitude relativistic Alfv\'{e}n waves into the inner magnetospheres~\cite{YuanMagnetarQuake2020,Bransgrove2020QuakeAlfven,ChenStarvation2022}. As these nonlinear waves propagate outward, they can drive the system into turbulence and initiate a cascade toward kinetic scales~\cite{ThompsonStarvation1998,Li2019AlfvenTurb}.
Owing to the highly relativistic nature of the accelerated leptons, the resulting PKT is inevitably influenced by radiative cooling~\cite{zhdankin2020kinetic,Comisso2021,nattila2021radiative,mehlhaff2025radiative,bilbao2025radiative}.
Furthermore, when the electron local quantum strong-field parameter
reaches $\chi_e \sim 0.1$, the turbulence enters an extreme regime in which macroscopic plasma dynamics becomes coupled to microscopic quantum electrodynamic (QED) processes~\cite{di2012extremely,uzdensky2014plasma,fedotov2023advances}.
In this regime, radiation reaction no longer acts as a continuous cooling force but proceeds through discrete high-energy photon emissions accompanied by substantial recoils and spin flips~\cite{Baier_1967,Sorbo_2017,Seipt_2018,li2019ultrarelativistic,geng2020spin,li2020polarized,seipt2023kinetic,qian2023parametric,qian2025fully}. 
Such QED processes have been shown to generate net spin polarization (SP) in pair cascades within pulsar polar caps~\cite{song2024polarized} and anomalous $\gamma$-ray polarization during magnetic reconnection~\cite{gong2025spin}.}
However, whether the macroscopic SP can spontaneously emerge in ultrarelativistic radiative PKT, and to what extent the spin signatures encode information about the underlying turbulence dynamics in strongly magnetized environments, remain open questions.

In this Letter, we carry out a spin-resolved study of the strongly magnetized electron-positron PKT in the radiation reaction dominated realm.
PIC simulations show that leptons are rapidly accelerated by inductive electric fields during the initial nonequilibrium stage, followed by intense radiative spin flips as the turbulence develops, leading to pronounced net SP.
\black{The spatiotemporal evolution of electron SP exhibits distinct characteristics between density-dominated (DD) and electromagnetic (EM) turbulence, with the latter being known also as charge-starved turbulence~\cite{blaes1989neutron_starvation,ThompsonStarvation1998,KumarStarvation2020,ChenStarvation2022,nattila2022heating,BoldyrevStarvation2025}. The detailed investigation of the spin accumulation and the spin depolarization processes led us to further classification of the EM-regime }
into dissipative and nondissipative states depending on the degree of radiation-induced turbulent dissipation. 
\black{In the DD regime, electrons exhibit only moderate SP and undergo nearly isotropic depolarization. In contrast, the EM regime supports much higher SP levels, which, after saturation, either decay anisotropically or remain persistent depending on the subtle dissipative or nondissipative states}. 
Therefore, the radiative SP, naturally emerging in ultrarelativistic magnetized PKT, introduces an additional degree of freedom that `probes' otherwise inaccessible turbulent regimes in extremely magnetized radiative plasmas.

\begin{figure}
\includegraphics[width=0.48\textwidth]{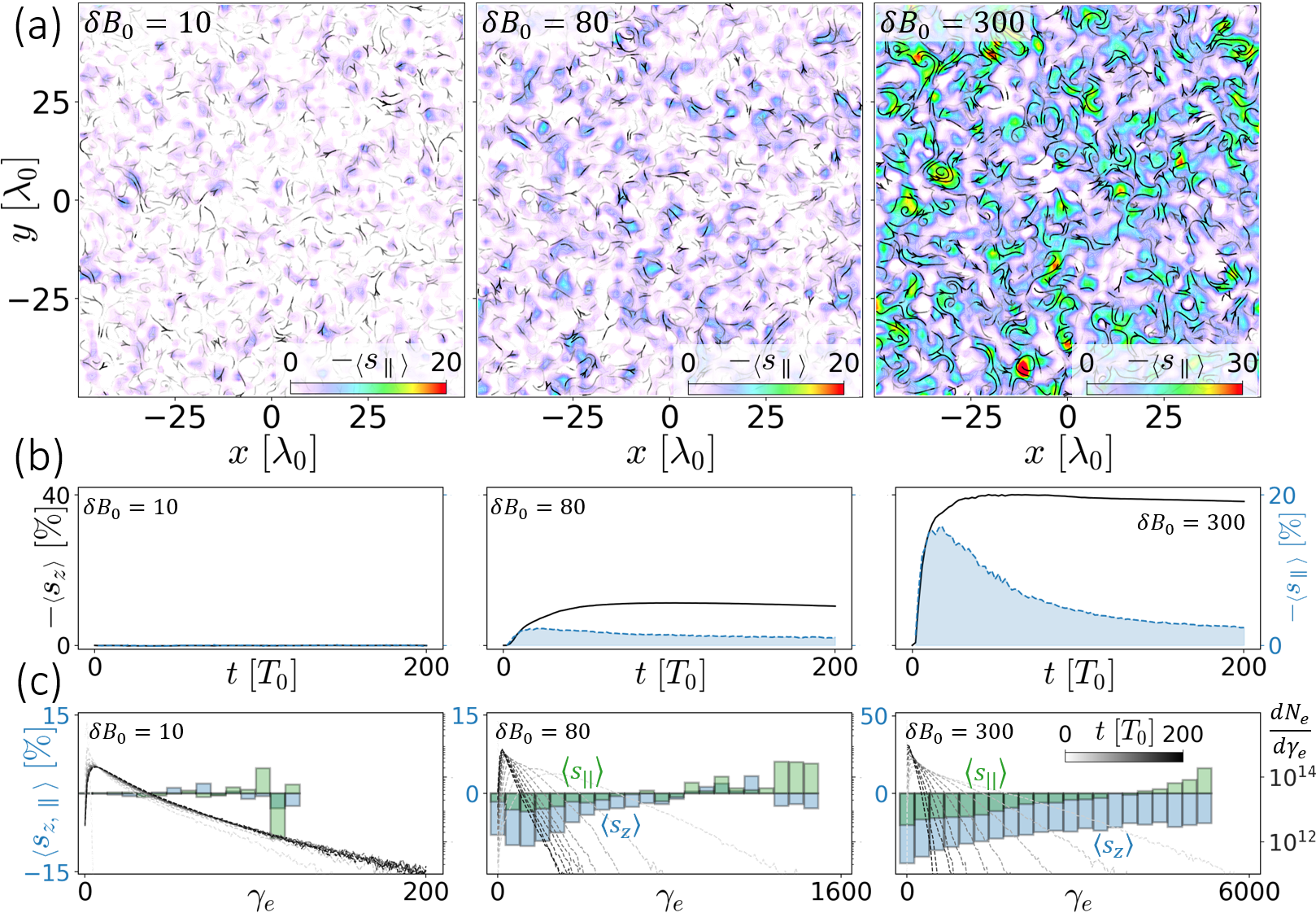}
\caption{\black{Spin-resolved turbulent dynamics for $n_e=n_0$. (a) Electron SP ratio $\left<s_{\parallel}\right>$ (expressed in percentage) at $t=20T_0$. The streamlines in (a) mark the $\left<s_{\parallel}\right>$ direction.
(b) Time evolution of $\left<s_z\right>$ and $\left<s_\parallel\right>$.
(c) The dependence of $\left<s_z\right>$ and $\left<s_\parallel\right>$ on $\gamma_e$ at $t=20T_0$, alongside the energy spectra $dN_e/d\gamma_e$.}
}
\label{fig:fig1_general_turb}
\end{figure}

The simulations are performed \black{by using the EPOCH code~\cite{arber2015contemporary}} with a spin-resolved implementation of strong-field QED processes~\cite{SM}.
While in the PIC code the photon emissions and the spin-flips are simulated statistically numerically~\cite{SM}, the electron (positron) spin dynamics can be followed and confirmed using the  equation~\cite{SM}:
\begin{eqnarray}\label{eq:spin_eq}
\frac{d\bm{s}}{dt}\approx\bm{s}\times\bm{\Omega}_e -\mathcal{A}(\chi_e)\hat{\bm{e}}_2  -\mathcal{B}(\chi_e)\bm{s}.
\end{eqnarray}
The first term in Eq.~\eqref{eq:spin_eq} describes the spin precession according to the Thomas-Bargmann-Michel-Telegdi equation~\cite{thomas1927kinematics,bargmann1959precession} where the precession frequency is \black{{\footnotesize $\bm{\Omega}_e= \frac{e}{2m_e}\left\{g\left(\frac{\bm{B}}{\gamma_e}-\frac{1}{\gamma_e+1}\frac{\bm{v}}{c^2}\times\bm{E}\right)\\ -(g-2)\frac{\gamma_e}{\gamma_e+1}\left[\frac{\bm{v}}{c^2}\times(\bm{E}+\bm{v}\times\bm{B})\right]\right\}$} with $\gamma_e$ the electron Lorentz factor and $g\approx2.00232$ the gyromagnetic factor. 
The remaining terms account for radiative spin flips $d\bm{s}_{rad}/dt=-\mathcal{A}(\chi_e)\hat{\bm{e}}_2  -\mathcal{B}(\chi_e)\bm{s}$}, where {\small $ \mathcal{A}(\chi_e)= \sqrt{3}\alpha_f  m_ec^2 \chi_e\mathcal{A}^*(\chi_e)/(2\pi\hbar\gamma_e)$}  and {\small $ \mathcal{B}(\chi_e)= \sqrt{3}\alpha_f  m_ec^2 \chi_e\mathcal{B}^*(\chi_e)/(2\pi\hbar\gamma_e)$} 
with $\mathcal{A}^*(\chi_e)\approx0.18\chi_e$ and $ \mathcal{B}^*(\chi_e)\sim \mathcal{A}^*(\chi_e)$ at $\chi_e<0.4$~\cite{SM}. 
Here, \black{$\alpha_f\approx 1/137$ is the fine-structure constant,} and $\hat{\bm{e}}_2$ is the magnetic field unit vector in the electron rest frame.
%

In our simulations of PKT, electrons and positrons with equal densities $n_e=n_p$ are initially homogeneous and thermalized with temperature $T_e = 0.3m_ec^2$. The two-dimensional simulation periodic domain of  $100 \lambda_0 \,\times\,100 \lambda_0$ is resolved 
by $2000\times2000$ cells, where $\lambda_0=2\pi c/\omega_0$ and $\omega_0=(n_0 e^2/ \varepsilon_0 m_e)^{1/2}$ with the speed of light $c$, electron mass (charge) $m_e$ ($e$), fiducial plasma density $n_0$, and vacuum permittivity $\varepsilon_0$.
A uniform guide magnetic field $\left< \mathbf{B} \right> = B_0 \hat{z}$ is imposed, superposed with transverse fluctuations $\delta B_{x,y}$, where $\left<...\right>$ denotes averages. 
The turbulence is initialized by uncorrelated large-scale modes~\cite{comisso2018particle,SM}:
$\delta B_x = \sum_{m,n} \beta_{mn} \, n \sin \Phi_{mn} \cos \Psi_{mn}$ and 
$\delta B_y = -\sum_{m,n} \beta_{mn} \, m \cos \Phi_{mn} \sin \Psi_{mn}$
where $\Phi_{mn}=k_m x + \phi_{mn}$, $\Psi_{mn}=k_n y + \psi_{mn}$, and $k_{m,n} = 2\pi (m,n)/L$ with mode indices $m,n \in \{1,\ldots,N\}$, mode number $N=16$, and $L=100\lambda_0$.
The phases $\phi_{mn}$ and $\psi_{mn}$ are random, and the amplitudes $\beta_{mn} = 2\delta B_{0}/[N(m^2+n^2)^{1/2}]$ ensure equal magnetic energy per mode, yielding a spectrum plateau near $k_N = 2\pi /\delta l$ with $\delta l\equiv L/N$. 
\black{The electron density $n_e$ and magnetic field $\delta B_0$ are normalized to $n_0\equiv 10^{21}\mathrm{cm}^{-3} $ and $m_e\omega_0/|e|\approx 10^8\mathrm{G}$, respectively. The primary example, $n_e=n_0$ and 
$\delta B_0=300 \approx 3\times10^{10}\mathrm{G}$, is relevant to a situation of inner magnetar magnetospheres, corresponding} 
to a cold magnetization $\sigma_{c}= (|e|\delta B_0/m_e\omega_0)^2/(n_e/n_0)\sim 10^5$.
%
This plasma can attain a quantum invariant $\chi_e\equiv (e\hbar/m_e^3c^4)|F_{\mu\nu}p^\mu|\gtrsim 0.1$ with $F_{\mu\nu}$ ($p^\mu$) the field (momentum) four-tensor, where radiative energy losses and spin-flip processes significantly influence the dynamics.
%

\black{We are interested in the strong turbulence scenario with magnetic lines undergoing violent twisting where the fluctuating field is comparable to that of the background, i.e. $\delta B_{0}/B_0 \sim 1$~\cite{zhdankin2017kinetic,comisso2018particle,Zhdankin2019,gorbunov2025leaking}. The gradual emergence of turbulence is observed with increasing the magnetic fluctuations.  In our study we monitor SP, see Fig.~\ref{fig:fig1_general_turb}, to find specific correlations with PKT features. The latter shows}
the results for three cases with $n_e=n_0$ and $\delta B_0=\left\{10,80,300\right\} $.
When the magnetic fluctuation increases from $\delta B_0=10$ to $300$,
a pronounced in-plane SP $s_\parallel\equiv (s_xB_x+s_yB_y)/(B_x^2+B_y^2)^{1/2}$ emerges with a local vortex-structured distribution during the initial nonequilibrium phase of turbulence development [Fig.~\ref{fig:fig1_general_turb}(a)]. 
For $\delta B_0=300$, the evolution of SP demonstrates that $s_z$ reaches saturation at $\left< s_z\right>\approx -40.1\%$ around $t= 20T_0$ \black{($T_0=\lambda_0/c$)} and subsequently persists throughout the simulation [Fig.~\ref{fig:fig1_general_turb}(b)], whereas $s_\parallel$ decays quickly after the saturation of $\left< s_\parallel\right>\approx -15.8\%$ at $t\approx 18T_0$. 
The result of $\left< s_\parallel\right> <0$ means that the in-plane spin components $\left<s_x\right>$ and $\left<s_y\right>$ preferentially align antiparallel to the local field $B_{xy}$. The energy-resolved SP of $\left< s_{z,\parallel}\right>$ vs $\gamma_e$  shows that lower-energy electrons attain higher polarization with $\left< s_{z,\parallel}\right> <0$ [Fig.~\ref{fig:fig1_general_turb}(c)], consistent with cumulative radiative spin flips contributed by $-\mathcal{A}(\chi_e)\hat{\bm{e}}_2$ in Eq.\eqref{eq:spin_eq}. 
The transient high SP ratio is attributed to the energetic electrons at the beginning, where turbulent magnetic energy is rapidly converted into particle kinetic energy during an initial acceleration dominated by the inductive field $E_z$ arising from $\nabla\times \bm{E}=-\partial\bm{B}/\partial t$. 
The simulation results show the maximum electron energy following a scaling $\gamma_e^{\max}\sim 10 \delta B_0$
 at $n_e\lesssim n_0$ and the corresponding invariant $\chi_e\sim 10|e|\delta B_0^2/B_{\text{c}}$,
where $B_{\text{c}}= m_{e}^{2}c^{2}/|e|\hbar \simeq 4.41\times 10^{13}\text{G}$ is the QED critical field.
As the radiative SP is significant at $\chi_e\gtrsim0.1$, the criterion of entering the SP-dominated PKT is estimated as $\delta B_0\gtrsim 10^2$,
confirmed by the SP $\left< s_{z,\parallel}\right>$ and its energy dependence [Fig.~\ref{fig:fig1_general_turb}(b)(c)].


\begin{figure}[t]
\includegraphics[width=0.99\columnwidth]{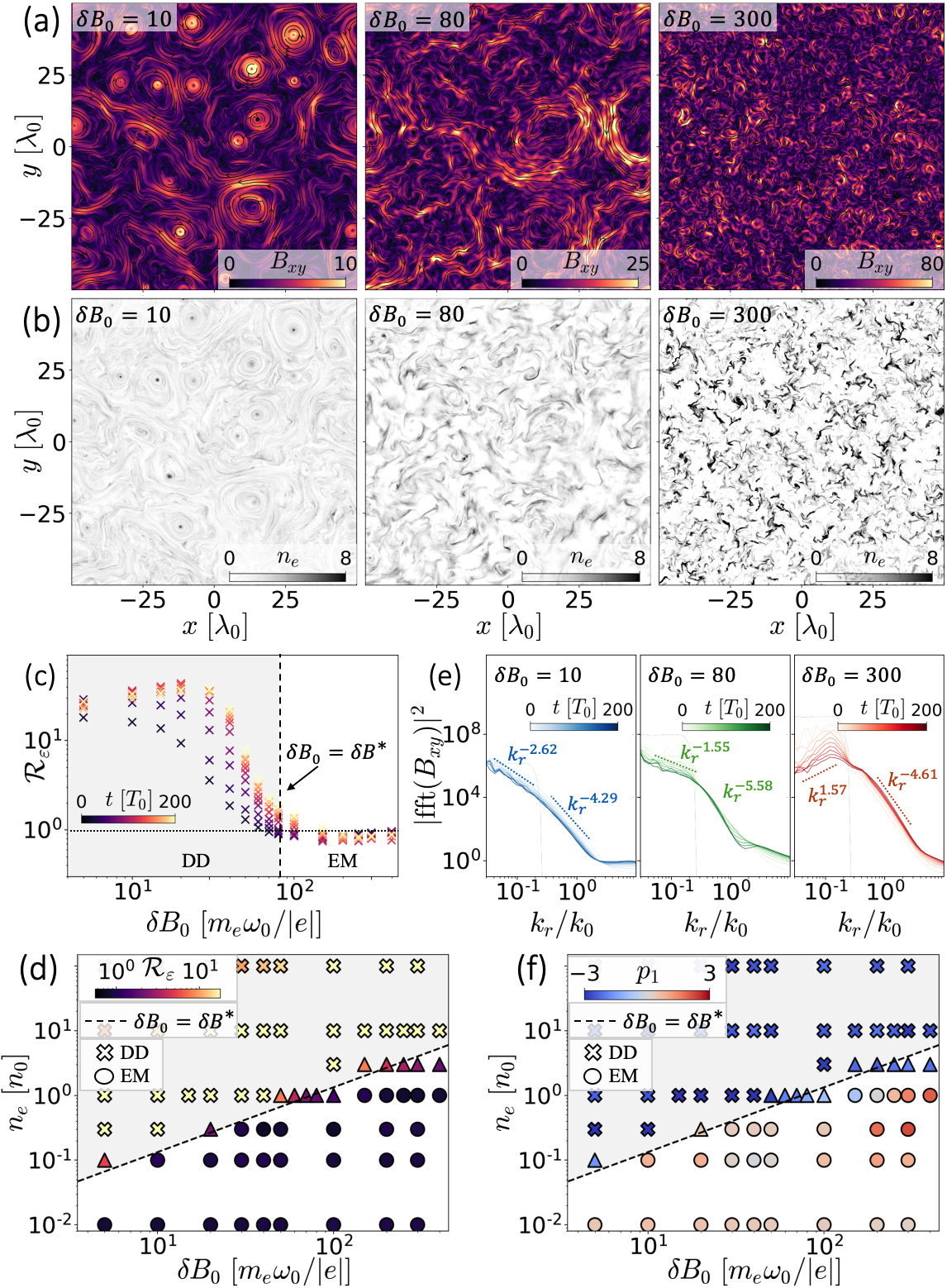}
\caption{\black{(a) Magnetic field $B_{xy}$ and (b) electron density $n_e$ at $t=150T_0$. The streamlines in (a) mark the $B_{xy}$ direction.
(c) Dependence of $\mathcal{R}_\varepsilon$ on $\delta B_0$; the vertical line denotes the threshold of Eq.\eqref{eq:th_DD_to_EM}.
(d) Dependence of $\mathcal{R}_\varepsilon$ on $\delta B_0$ and $n_e$.
(e) Magnetic energy spectra $\mathrm{|fft}(B_{xy})|^2$ vs $k_r$. 
(f) Dependence of $p_1$ on $\delta B_0$ and $n_e$.
In (a)(b)(c)(e), the density is $n_e=n_0$. 
In (d),(f)  markers indicate the  simulation in the EM (circle), transition (triangle), and DD regimes (cross).
}}
\label{fig:fig2_fft_R_spin}
\end{figure}

\black{With SP behavior at hand [Fig.~\ref{fig:fig1_general_turb}], now we inspect the magnetic fields $B_{xy}$ and electron densities $n_e$ in the same parameter range, exhibiting distinct PKT [Fig.~\ref{fig:fig2_fft_R_spin}(a)(b)].}
For $\delta B_0=10$, magnetic fluctuations undergo coalescence, forming discrete large-scale swirling structures with turbulent vortices $B_{xy}$. 
This magnetic fluctuation is primarily contributed by the current density $j_z\sim|e|n_ev_z$, naturally termed as density dominated (DD). The particle acceleration and radiative cooling in the DD regime were comprehensively  studied~\cite{zhdankin2017kinetic,comisso2018particle,Zhdankin2019,Comisso2021,nattila2022heating,ComissoAPJL2022,nattila2024radiative}.
%
For $\delta B_0=300 $, the initialized magnetic strength of $B_{xy}\sim\delta B_0$ is so high that the quasi-static field comparable to $\delta B_0$ is sustained mostly by the displacement current $\varepsilon_0\partial\bm{E}/\partial t$, rather than by the current density $j_z$,
and thus the turbulence evolves into an EM state, \black{analogous to charge-starvation effects previously discussed in highly magnetized astrophysical scenarios ~\cite{blaes1989neutron_starvation,ThompsonStarvation1998,KumarStarvation2020,nattila2022heating,ChenStarvation2022,BoldyrevStarvation2025}.}
The EM-regime is characterized by the near equipartition between electric and magnetic energy densities, 
 $\varepsilon_0E_z^2\sim B_{xy}^2/\mu_0$, thus the turbulence regime can be classified through ratio $\mathcal{R}_\varepsilon\equiv \left< B_{xy}^2\right>/\left<\mu_0\varepsilon_0E_z^2\right> $. Thus, the PKT is in the DD regime when $\mathcal{R}_\varepsilon \gg 1$, while at $\mathcal{R}_\varepsilon \to 1$ it enters in the EM-regime [Fig.~\ref{fig:fig2_fft_R_spin}(c)].
The transition threshold from the DD to EM regime $\mathcal{R}_\varepsilon > 1$ can be represented as~\cite{SM}:
\begin{equation}\label{eq:th_DD_to_EM}
\delta B_0 > \delta B^* \sim 2\pi^2 m_e\omega_0 L n_e/ (|e|\lambda_0 n_0 N).
\end{equation}
This threshold is proved by the parameter scan simulations of the dependence of $\mathcal{R}_\varepsilon$ on $n_e$ and $\delta B_0$ [Fig.~\ref{fig:fig2_fft_R_spin}(d)]. \black{Comparison of Figs.~\ref{fig:fig1_general_turb}(a) and \ref{fig:fig2_fft_R_spin}(a) shows that the high SP  is correlated with the EM regime. We will see below that SP transient dynamics allows for the EM sub-regimes' classification.}

\black{The typical feature of turbulence is the vortex spatial scale. 
To quantify this property in PKT, we investigate the spatial Fourier transform of the magnetic energy spectra,} see $|\mathrm{fft}(B_{xy})|^2$ as a function of wavenumber $k_r \equiv(k_x^2+k_y^2)^{1/2}$ in Fig.~\ref{fig:fig2_fft_R_spin}(e).
For the DD regime with 
$\delta B_0=10$, the spectrum follows a two-stage decaying power index \black{with $p_1\approx -2.62$ at $k_r \sim0.1d_e^{-1} \ (d_e \equiv c/\omega_0)$, similar to the prediction of the intermittency model  $E(k_r) \sim k_r^{-8/3}$~\cite{Boldyrev2012,WanIntermittent2012,Zhou2023}. For the $k_r \sim d_e^{-1}$, the spectrum steepens to $p_2\approx -4.29$, similar to the scalings of the relativistic magnetized turbulence~\cite{ComissoAPJ2019,nattila2021radiative}. }

\black{In the EM regime 
($\delta B_0=300$ and $n_e = n_0$), a surprising feature is observed in the spectrum: a positive slope $p_1 \approx 1.57$ at $k_r \lesssim 0.1d_e^{-1}$ [Fig.~\ref{fig:fig2_fft_R_spin}(e)], i.e., the strong radiation damping leads to substantial dissipation of large-scale magnetic fluctuations $k_r\rightarrow 0$, while} the spectrum maintains a negative slope $p_2 \approx -4.61$ at higher frequency, similar to the cascade dissipative behavior observed in the lower-$\delta B_0$ cases. \black{An intuitive explanation for the power index $ p_1\sim1.57$ induced by strong radiative dissipation is given in the End Matter, and it is demonstrated that the scaling $E(k_r) \sim k_r^2$ holds in the strong synchrotron cooling regime.
Note that $p_1>0$ is typical and ubiquitous for the EM regime, as confirmed by the dependence of $p_1$ on $n_e$ and $\delta B_0$ [Fig.~\ref{fig:fig2_fft_R_spin}(f)], where the cases with $p_1\sim 2$ is mostly correlated with the radiation-induced dissipative EM regime near the parameters of $n_e=n_1$ and $\delta B_0=300 $.}

\begin{figure}
\includegraphics[width=0.95\columnwidth]{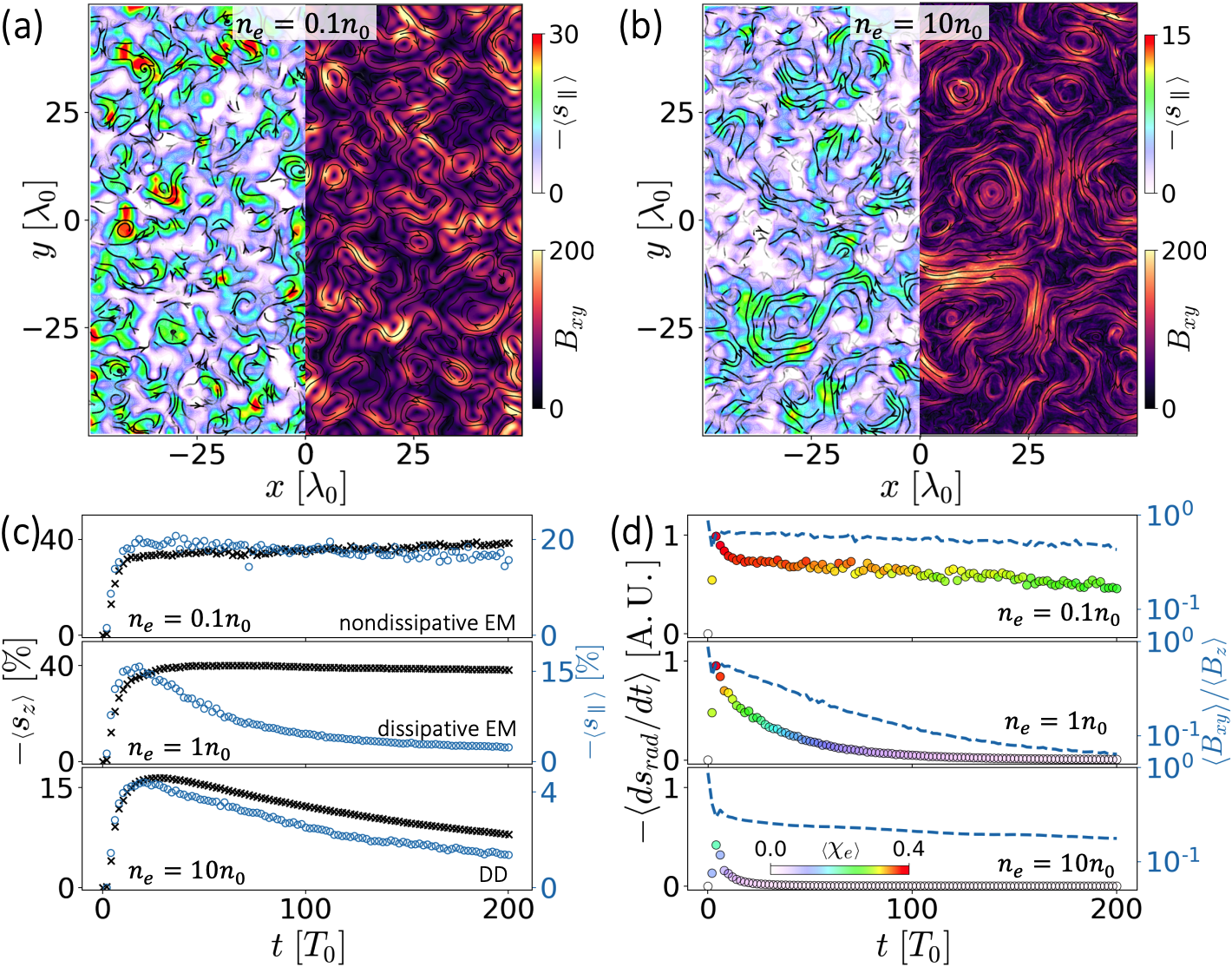}
\caption{\black{Magnetic field $B_{xy}$ and SP for $\delta B_0=300$: 
Distributions of $\left<s_\parallel\right>$ at $t =20T_0$ and $B_{xy}$ at $t=150 T_0$ for (a) $n_e=0.1n_0$, 
b)  for $n_e=10n_0$;
(c) Time evolution of  $\left<s_z\right>$ and  $\left<s_\parallel \right>$;
(d) Time evolution of 
$\left<d s_{rad}/dt\right>$ (cycles, with  rainbow color denoting $\left<\chi_e\right>$) and $\left<B_{xy}\right>/\left<B_z\right>$ (dashed line).
}}
\label{fig:fig3_s_relax}
\end{figure}

\black{We further explore the transient time-dependent SP signal in the EM-regime as an information source for PKT, in particular, analyzing the spin accumulation and the spin depolarization at $\delta B_0=300  $ and $n_e = \left\{0.1,1,10\right\}n_0$, see  Fig.~\ref{fig:fig3_s_relax}. The time behavior of $\left<s_{\parallel,z}\right>$  in the DD-regime $n_e =10 n_0$ is different from that of the EM-regime as expected [Fig.~\ref{fig:fig3_s_relax}(c)]. However, we observe two distinct SP behaviors in the EM-regime for $n_e = 0.1n_0$ and $n_e =1 n_0$. In the EM case $n_e = 0.1n_0$,}  both  $\left<s_\parallel\right>$ and $\left<s_z\right>$  remain elevated throughout the interaction, with $\left<s_z\right>$ even exhibiting a slight enhancement [Fig.~\ref{fig:fig3_s_relax}(c)]. The explanation is that in-plane magnetic fields and electrons maintain high energy levels, allowing a high quantum invariant $\left<\chi_e\right>$ and radiative spin flips  ($\left< s_{rad}/dt=-\mathcal{A}(\chi_e)  -\mathcal{B}(\chi_e)(\bm{s}\cdot\hat{\bm{e}}_2) \right>$)  to persist [Fig.~\ref{fig:fig3_s_relax}(d)], where the turbulence stays at a high-SP state. 

For the  EM case $n_e = n_0$, the relaxation of $\left<s_\parallel\right>$ is significant, while $\left<s_z\right>$ is largely preserved [Fig.~\ref{fig:fig3_s_relax}(c)], implying an asymmetric magnetic energy partition, $B_z \gg B_{xy}$, favoring depolarization via spin precession in the $(x,y)$ plane. 
We trace the origin of the two EM sub-regimes in the dissipation channel of the turbulent energy. For  $n_e = 0.1n_0$, the magnetic spectra with a reduced positive power index of $ p_1 \sim 0.74$ indicate that although radiation-induced dissipation still exists, it gets suppressed due to the lower electron population. 
The distribution of SP ratio $\left<s_\parallel\right>$ [Fig.~\ref{fig:fig3_s_relax}(a)] exhibits small-scale vortices following the magnetic field $B_{xy}$ with an unweakened strength.
Therefore, the case of $n_e = 0.1n_0$ is regarded as a non-dissipative EM regime compared to the dissipative EM regime for the case of $n_e=n_0$. 
In the case of $n_e = n_0$, the anisotropic depolarization occurs after the $ B_{xy}$  energy has been efficiently transferred to leptons and then radiated away [Fig.~\ref{fig:fig3_s_relax}(d)], and the system enters a dissipated EM turbulence without further radiative spin flips. 
For  $n_e = 10n_0$ ($\delta B_0<\delta B^*$) belonging to the DD regime, the magnetic spectra show a power index $p_1\approx-2.49<0$ corresponding to the normal turbulence cascade direction and the SP $\left<s_\parallel\right>$ with large-scale swirling structures aligned with the local magnetic vortices [Fig.~\ref{fig:fig3_s_relax}(b)]. Here,  both $\left<s_\parallel\right>$  and $\left<s_z\right>$ undergo apparent relaxation [Fig.~\ref{fig:fig3_s_relax}(c)]. 
This decay results from no further radiative spin-flip events and nearly isotropic depolarization. The latter is due to the spin precession induced by both $B_{xy}$ and $B_z$ since the radiation-induced energy dissipation of magnetic field $B_{xy}$ is not as strong as that in the dissipative EM case [Fig.~\ref{fig:fig3_s_relax}(d)].

\begin{figure}[t]
\includegraphics[width=0.98\columnwidth]{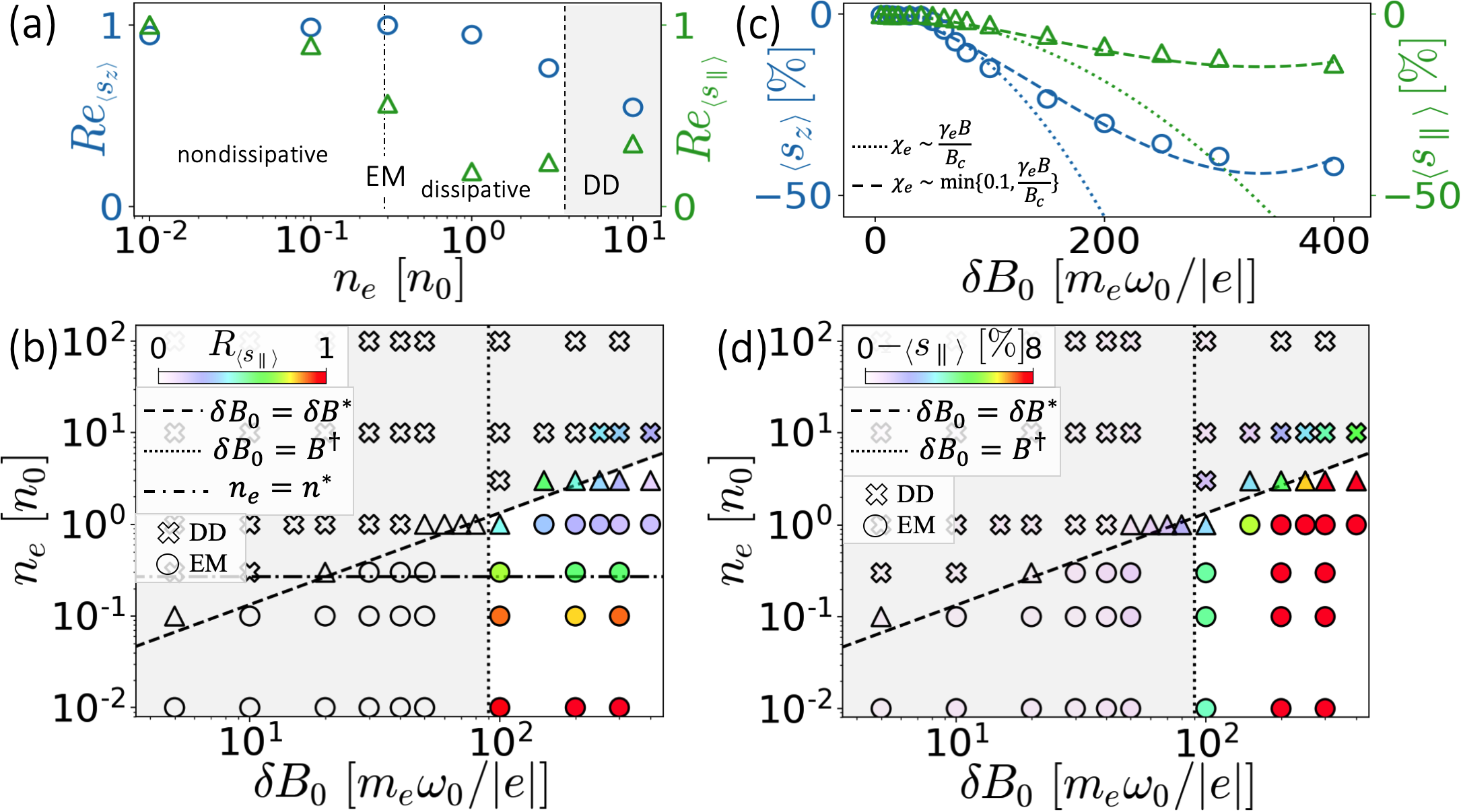}
\caption{\black{Parameter scans delimiting the dissipative EM-regime:}  
(a) Dependence of the SP relaxation $Re\left<s_{z,\parallel}\right>$ on $n_e$ for $\delta B_0=300 $.
(b) Dependence of $Re\left<s_\parallel\right>$ on $\delta B_0$ and $n_e$ for the cases with $\left<s_\parallel\right><-2\%$.
(c) Dependence of $\left<s_{z,\parallel}\right>$ on $\delta B_0$ for $n_e=n_0$. 
(d) Dependence of $\left<s_\parallel\right>$ on $\delta B_0$ and $n_e$.
Here, lines represent the analytical prediction. The markers indicate the regimes as in Fig.~2.}
\label{fig:fig4_scan}
\end{figure}

\black{Thus, the separation of the EM-regime into the nondissipative and dissipative sub-regimes is distinguished by the role of depolarization.}
To quantify the depolarization, the SP relaxation ratio $Re\left<s\right>$ is defined as the ratio of $|\left<s\right>|$ at the final time to its maximum. 
The dependence of $Re\left<s_{z,\parallel}\right>$ on $n_e$ for $\delta B_0=300$ shows distinct isotropic and anisotropic depolarization among the DD and dissipative EM regimes [Fig.~\ref{fig:fig4_scan}(a)]. 
The criterion for entering the dissipative EM state is approximated as $n_e\gtrsim n^*\sim10^{-2}\rho_1 |e| \delta B_0 n_0/(m_e\omega_0 \gamma_e \chi_e)\sim0.3 n_0 $ through $P_{rad}\delta t \gtrsim \rho_1 \delta B_0^2/2\mu_0$ as proved in Fig.~\ref{fig:fig4_scan}(b), where $P_{rad}\sim |e|c^2\gamma_e\chi_e\delta B_0n_e$ is the radiation power, $\delta t\sim10^2T_0$ the interaction time, $ \delta B_0^2/2\mu_0$ the EM energy of the system, $\rho_1\sim1$ a constant coefficient, and $\gamma_e\propto|e|\delta B_0L/m_e\omega_0\lambda_0$ is utilized.
\black{The parameter scans delimiting the dissipative EM-regime are presented in Fig.~\ref{fig:fig4_scan}(a)(b).}
Besides, the accumulated SP ratio is estimated as $\left<s_{z,\parallel}\right>\propto -\sqrt{3}\alpha_f m_ec^2\chi_e^2/2\pi\hbar\gamma_e $ by using $d s_\mathrm{rad}/dt \approx-\sqrt{3}\alpha_f m_ec^2 \delta B_0 \mathcal{A}^*(\chi_e)/2\pi\hbar B_\text{c}$ in Eq.\eqref{eq:spin_eq}, where $\mathcal{A}^*(\chi_e)\approx0.18\chi_e$ and $\chi_e\sim \mathrm{min}\{0.1,\gamma_e \delta B_0/B_{\text{c}}\}$ is employed~\cite{SM}. This scaling agrees well with the simulation results [Fig.~\ref{fig:fig4_scan}(c)].
%
The parameter scan of $\left<s_\parallel\right>$ indicates that a high SP ratio is achieved when $\delta B_0>\mathrm{max}\{B^\dagger,\,\delta B^*\}$ with $B^\dagger \sim 10^2 $ derived from significant spin flips with $\mathcal{A}(\chi_e)\gtrsim 10^{-2}T_0^{-1}$ and $\delta B_0\gtrsim\delta B^*$ the criterion of the EM regime [Fig.~\ref{fig:fig4_scan}(d)].

In conclusion, we demonstrate a dynamical SP state in ultrarelativistic PKT. The turbulence with a high SP value is associated with the EM regime, exhibiting \black{a nontrivial positive power-law index in the magnetic spectra}. The isotropic, anisotropic, and absent depolarization are correlated with the DD, dissipative EM, and non-dissipative EM regimes of PKT, respectively. 
\black{These findings suggest the intrinsic existence of spin-polarized turbulence in magnetar giant flares triggered by starquakes~\cite{ThompsonQuake2001}, as well as in magnetospheres perturbed by nonlinear interactions of Alfv\'{e}n waves~\cite{Li2019AlfvenTurb}, potentially resulting in anomalous $\gamma$-ray polarization~\cite{gong2025spin} or providing an interpretation of the nontrivial polarization angles of emission from gamma-ray bursts~\cite{zhang2019detailed}}.
The dissipative EM regime may constitute a key element of highly magnetized turbulence in soft $\gamma$-ray repeaters~\cite{kouveliotou1993recurrent,Thompson1995MNRAS} and energetic pulsars~\cite{hess2023discovery,hakobyan2023magnetic}. 
With the advent of high-intensity beam facilities, SP turbulence might become accessible in the near future~\cite{gonoskov2022charged,chen2023perspectives,arrowsmith2024laboratory,mercuri2025growth,pouyez2025kinetic,reichwein2025plasma}, opening new avenues for understanding PKT across diverse scenarios.



\begin{acknowledgments}
We acknowledges the support from the Strategic Priority Research Program of the Chinese Academy of Sciences (Grant No. XDB1550100), the National Key R\&D Program of China (2025YFF0515103), the CAS Project for Young Scientists in Basic Research (Grant No. YSBR-141), the NNSFC (Grant No. 12447101), and the HPC Cluster of ITP-CAS.
The original version of code EPOCH is funded by the UK EPSRC grants EP/G054950/1, EP/G056803/1, EP/G055165/1 and EP/M022463/1.
The authors thank Matteo Tamburini, Luca Comisso, Vladimir Zhdankin, and Muni Zhou for fruitful discussions.
We also thank the anonymous referees for their useful comments and suggestions. 

Data availability---The data supporting this article are openly available~\cite{Data}, embargo periods may apply.
\end{acknowledgments}

\bibliography{aa}

\begin{thebibliography}{104}%
\makeatletter
\providecommand \@ifxundefined [1]{%
 \@ifx{#1\undefined}
}%
\providecommand \@ifnum [1]{%
 \ifnum #1\expandafter \@firstoftwo
 \else \expandafter \@secondoftwo
 \fi
}%
\providecommand \@ifx [1]{%
 \ifx #1\expandafter \@firstoftwo
 \else \expandafter \@secondoftwo
 \fi
}%
\providecommand \natexlab [1]{#1}%
\providecommand \enquote  [1]{``#1''}%
\providecommand \bibnamefont  [1]{#1}%
\providecommand \bibfnamefont [1]{#1}%
\providecommand \citenamefont [1]{#1}%
\providecommand \href@noop [0]{\@secondoftwo}%
\providecommand \href [0]{\begingroup \@sanitize@url \@href}%
\providecommand \@href[1]{\@@startlink{#1}\@@href}%
\providecommand \@@href[1]{\endgroup#1\@@endlink}%
\providecommand \@sanitize@url [0]{\catcode `\\12\catcode `\$12\catcode
  `\&12\catcode `\#12\catcode `\^12\catcode `\_12\catcode `\%12\relax}%
\providecommand \@@startlink[1]{}%
\providecommand \@@endlink[0]{}%
\providecommand \url  [0]{\begingroup\@sanitize@url \@url }%
\providecommand \@url [1]{\endgroup\@href {#1}{\urlprefix }}%
\providecommand \urlprefix  [0]{URL }%
\providecommand \Eprint [0]{\href }%
\providecommand \doibase [0]{https://doi.org/}%
\providecommand \selectlanguage [0]{\@gobble}%
\providecommand \bibinfo  [0]{\@secondoftwo}%
\providecommand \bibfield  [0]{\@secondoftwo}%
\providecommand \translation [1]{[#1]}%
\providecommand \BibitemOpen [0]{}%
\providecommand \bibitemStop [0]{}%
\providecommand \bibitemNoStop [0]{.\EOS\space}%
\providecommand \EOS [0]{\spacefactor3000\relax}%
\providecommand \BibitemShut  [1]{\csname bibitem#1\endcsname}%
\let\auto@bib@innerbib\@empty
\bibitem [{\citenamefont {Gary}(1993)}]{gary1993theory}%
  \BibitemOpen
  \bibfield  {author} {\bibinfo {author} {\bibfnamefont {S.~P.}\ \bibnamefont
  {Gary}},\ }\href@noop {} {\emph {\bibinfo {title} {Theory of space plasma
  microinstabilities}}},\ \bibinfo {number} {7}\ (\bibinfo  {publisher}
  {Cambridge university press},\ \bibinfo {year} {1993})\BibitemShut {NoStop}%
\bibitem [{\citenamefont {Horton}(1999)}]{horton1999drift}%
  \BibitemOpen
  \bibfield  {author} {\bibinfo {author} {\bibfnamefont {W.}~\bibnamefont
  {Horton}},\ }\bibfield  {title} {\bibinfo {title} {Drift waves and
  transport},\ }\href@noop {} {\bibfield  {journal} {\bibinfo  {journal}
  {Reviews of Modern Physics}\ }\textbf {\bibinfo {volume} {71}},\ \bibinfo
  {pages} {735} (\bibinfo {year} {1999})}\BibitemShut {NoStop}%
\bibitem [{\citenamefont {Howes}\ \emph {et~al.}(2008)\citenamefont {Howes}
  \emph {et~al.}}]{howes2008kinetic}%
  \BibitemOpen
  \bibfield  {author} {\bibinfo {author} {\bibfnamefont {G.}~\bibnamefont
  {Howes}} \emph {et~al.},\ }\bibfield  {title} {\bibinfo {title} {Kinetic
  simulations of magnetized turbulence in astrophysical plasmas},\ }\href@noop
  {} {\bibfield  {journal} {\bibinfo  {journal} {Physical Review Letters}\
  }\textbf {\bibinfo {volume} {100}},\ \bibinfo {pages} {065004} (\bibinfo
  {year} {2008})}\BibitemShut {NoStop}%
\bibitem [{\citenamefont {Schekochihin}\ \emph {et~al.}(2009)\citenamefont
  {Schekochihin} \emph {et~al.}}]{schekochihin2009astrophysical}%
  \BibitemOpen
  \bibfield  {author} {\bibinfo {author} {\bibfnamefont {A.}~\bibnamefont
  {Schekochihin}} \emph {et~al.},\ }\bibfield  {title} {\bibinfo {title}
  {Astrophysical gyrokinetics: kinetic and fluid turbulent cascades in
  magnetized weakly collisional plasmas},\ }\href@noop {} {\bibfield  {journal}
  {\bibinfo  {journal} {The Astrophysical Journal Supplement Series}\ }\textbf
  {\bibinfo {volume} {182}},\ \bibinfo {pages} {310} (\bibinfo {year}
  {2009})}\BibitemShut {NoStop}%
\bibitem [{\citenamefont {Servidio}\ \emph {et~al.}(2012)\citenamefont
  {Servidio} \emph {et~al.}}]{servidio2012local}%
  \BibitemOpen
  \bibfield  {author} {\bibinfo {author} {\bibfnamefont {S.}~\bibnamefont
  {Servidio}} \emph {et~al.},\ }\bibfield  {title} {\bibinfo {title} {Local
  kinetic effects in two-dimensional plasma turbulence},\ }\href@noop {}
  {\bibfield  {journal} {\bibinfo  {journal} {Physical review letters}\
  }\textbf {\bibinfo {volume} {108}},\ \bibinfo {pages} {045001} (\bibinfo
  {year} {2012})}\BibitemShut {NoStop}%
\bibitem [{\citenamefont {Gro{\v{s}}elj}\ \emph {et~al.}(2019)\citenamefont
  {Gro{\v{s}}elj} \emph {et~al.}}]{grovselj2019kinetic}%
  \BibitemOpen
  \bibfield  {author} {\bibinfo {author} {\bibfnamefont {D.}~\bibnamefont
  {Gro{\v{s}}elj}} \emph {et~al.},\ }\bibfield  {title} {\bibinfo {title}
  {Kinetic turbulence in astrophysical plasmas: waves and/or structures?},\
  }\href@noop {} {\bibfield  {journal} {\bibinfo  {journal} {Physical Review
  X}\ }\textbf {\bibinfo {volume} {9}},\ \bibinfo {pages} {031037} (\bibinfo
  {year} {2019})}\BibitemShut {NoStop}%
\bibitem [{\citenamefont {Arzamasskiy}\ \emph {et~al.}(2023)\citenamefont
  {Arzamasskiy} \emph {et~al.}}]{arzamasskiy2023kinetic}%
  \BibitemOpen
  \bibfield  {author} {\bibinfo {author} {\bibfnamefont {L.}~\bibnamefont
  {Arzamasskiy}} \emph {et~al.},\ }\bibfield  {title} {\bibinfo {title}
  {Kinetic turbulence in collisionless high-$\beta$ plasmas},\ }\href@noop {}
  {\bibfield  {journal} {\bibinfo  {journal} {Physical Review X}\ }\textbf
  {\bibinfo {volume} {13}},\ \bibinfo {pages} {021014} (\bibinfo {year}
  {2023})}\BibitemShut {NoStop}%
\bibitem [{\citenamefont {Park}\ \emph {et~al.}(2025)\citenamefont {Park} \emph
  {et~al.}}]{park2025kinetic}%
  \BibitemOpen
  \bibfield  {author} {\bibinfo {author} {\bibfnamefont {J.~Y.}\ \bibnamefont
  {Park}} \emph {et~al.},\ }\bibfield  {title} {\bibinfo {title} {Kinetic
  turbulence drives mhd equilibrium change via 3d reconnection},\ }\href@noop
  {} {\bibfield  {journal} {\bibinfo  {journal} {Nature}\ }\textbf {\bibinfo
  {volume} {644}},\ \bibinfo {pages} {59} (\bibinfo {year} {2025})}\BibitemShut
  {NoStop}%
\bibitem [{\citenamefont {Blasi}(2013)}]{Blasi2013}%
  \BibitemOpen
  \bibfield  {author} {\bibinfo {author} {\bibfnamefont {P.}~\bibnamefont
  {Blasi}},\ }\bibfield  {title} {\bibinfo {title} {The origin of galactic
  cosmic rays},\ }\href@noop {} {\bibfield  {journal} {\bibinfo  {journal} {The
  Astronomy and Astrophysics Review}\ }\textbf {\bibinfo {volume} {21}},\
  \bibinfo {pages} {70} (\bibinfo {year} {2013})}\BibitemShut {NoStop}%
\bibitem [{\citenamefont {Matsumoto}\ \emph {et~al.}(2015)\citenamefont
  {Matsumoto}, \citenamefont {Amano}, \citenamefont {Kato},\ and\ \citenamefont
  {Hoshino}}]{matsumoto2015stochastic}%
  \BibitemOpen
  \bibfield  {author} {\bibinfo {author} {\bibfnamefont {Y.}~\bibnamefont
  {Matsumoto}}, \bibinfo {author} {\bibfnamefont {T.}~\bibnamefont {Amano}},
  \bibinfo {author} {\bibfnamefont {T.}~\bibnamefont {Kato}},\ and\ \bibinfo
  {author} {\bibfnamefont {M.}~\bibnamefont {Hoshino}},\ }\bibfield  {title}
  {\bibinfo {title} {Stochastic electron acceleration during spontaneous
  turbulent reconnection in a strong shock wave},\ }\href@noop {} {\bibfield
  {journal} {\bibinfo  {journal} {Science}\ }\textbf {\bibinfo {volume}
  {347}},\ \bibinfo {pages} {974} (\bibinfo {year} {2015})}\BibitemShut
  {NoStop}%
\bibitem [{\citenamefont {Zhdankin}\ \emph {et~al.}(2017)\citenamefont
  {Zhdankin}, \citenamefont {Werner}, \citenamefont {Uzdensky},\ and\
  \citenamefont {Begelman}}]{zhdankin2017kinetic}%
  \BibitemOpen
  \bibfield  {author} {\bibinfo {author} {\bibfnamefont {V.}~\bibnamefont
  {Zhdankin}}, \bibinfo {author} {\bibfnamefont {G.~R.}\ \bibnamefont
  {Werner}}, \bibinfo {author} {\bibfnamefont {D.~A.}\ \bibnamefont
  {Uzdensky}},\ and\ \bibinfo {author} {\bibfnamefont {M.~C.}\ \bibnamefont
  {Begelman}},\ }\bibfield  {title} {\bibinfo {title} {Kinetic turbulence in
  relativistic plasma: from thermal bath to nonthermal continuum},\ }\href@noop
  {} {\bibfield  {journal} {\bibinfo  {journal} {Physical Review Letters}\
  }\textbf {\bibinfo {volume} {118}},\ \bibinfo {pages} {055103} (\bibinfo
  {year} {2017})}\BibitemShut {NoStop}%
\bibitem [{\citenamefont {Comisso}\ and\ \citenamefont
  {Sironi}(2018)}]{comisso2018particle}%
  \BibitemOpen
  \bibfield  {author} {\bibinfo {author} {\bibfnamefont {L.}~\bibnamefont
  {Comisso}}\ and\ \bibinfo {author} {\bibfnamefont {L.}~\bibnamefont
  {Sironi}},\ }\bibfield  {title} {\bibinfo {title} {Particle acceleration in
  relativistic plasma turbulence},\ }\href@noop {} {\bibfield  {journal}
  {\bibinfo  {journal} {Physical review letters}\ }\textbf {\bibinfo {volume}
  {121}},\ \bibinfo {pages} {255101} (\bibinfo {year} {2018})}\BibitemShut
  {NoStop}%
\bibitem [{\citenamefont {Zhdankin}\ \emph {et~al.}(2019)\citenamefont
  {Zhdankin}, \citenamefont {Uzdensky}, \citenamefont {Werner},\ and\
  \citenamefont {Begelman}}]{Zhdankin2019}%
  \BibitemOpen
  \bibfield  {author} {\bibinfo {author} {\bibfnamefont {V.}~\bibnamefont
  {Zhdankin}}, \bibinfo {author} {\bibfnamefont {D.~A.}\ \bibnamefont
  {Uzdensky}}, \bibinfo {author} {\bibfnamefont {G.~R.}\ \bibnamefont
  {Werner}},\ and\ \bibinfo {author} {\bibfnamefont {M.~C.}\ \bibnamefont
  {Begelman}},\ }\bibfield  {title} {\bibinfo {title} {Electron and ion
  energization in relativistic plasma turbulence},\ }\href@noop {} {\bibfield
  {journal} {\bibinfo  {journal} {Physical Review Letters}\ }\textbf {\bibinfo
  {volume} {122}},\ \bibinfo {pages} {055101} (\bibinfo {year}
  {2019})}\BibitemShut {NoStop}%
\bibitem [{\citenamefont {Comisso}\ and\ \citenamefont
  {Sironi}(2022)}]{ComissoAPJL2022}%
  \BibitemOpen
  \bibfield  {author} {\bibinfo {author} {\bibfnamefont {L.}~\bibnamefont
  {Comisso}}\ and\ \bibinfo {author} {\bibfnamefont {L.}~\bibnamefont
  {Sironi}},\ }\bibfield  {title} {\bibinfo {title} {Ion and electron
  acceleration in fully kinetic plasma turbulence},\ }\href@noop {} {\bibfield
  {journal} {\bibinfo  {journal} {The Astrophysical Journal Letters}\ }\textbf
  {\bibinfo {volume} {936}},\ \bibinfo {pages} {L27} (\bibinfo {year}
  {2022})}\BibitemShut {NoStop}%
\bibitem [{\citenamefont {Vega}\ \emph {et~al.}(2022)\citenamefont {Vega} \emph
  {et~al.}}]{Vega2022}%
  \BibitemOpen
  \bibfield  {author} {\bibinfo {author} {\bibfnamefont {C.}~\bibnamefont
  {Vega}} \emph {et~al.},\ }\bibfield  {title} {\bibinfo {title} {Turbulence
  and particle acceleration in a relativistic plasma},\ }\href@noop {}
  {\bibfield  {journal} {\bibinfo  {journal} {The Astrophysical Journal
  Letters}\ }\textbf {\bibinfo {volume} {924}},\ \bibinfo {pages} {L19}
  (\bibinfo {year} {2022})}\BibitemShut {NoStop}%
\bibitem [{\citenamefont {Zhou}\ \emph
  {et~al.}(2023{\natexlab{a}})\citenamefont {Zhou}, \citenamefont {Liu},\ and\
  \citenamefont {Loureiro}}]{zhou2023electron}%
  \BibitemOpen
  \bibfield  {author} {\bibinfo {author} {\bibfnamefont {M.}~\bibnamefont
  {Zhou}}, \bibinfo {author} {\bibfnamefont {Z.}~\bibnamefont {Liu}},\ and\
  \bibinfo {author} {\bibfnamefont {N.~F.}\ \bibnamefont {Loureiro}},\
  }\bibfield  {title} {\bibinfo {title} {Electron heating in
  kinetic-alfv{\'e}n-wave turbulence},\ }\href@noop {} {\bibfield  {journal}
  {\bibinfo  {journal} {Proceedings of the National Academy of Sciences}\
  }\textbf {\bibinfo {volume} {120}},\ \bibinfo {pages} {e2220927120} (\bibinfo
  {year} {2023}{\natexlab{a}})}\BibitemShut {NoStop}%
\bibitem [{\citenamefont {Ewart}\ \emph {et~al.}(2025)\citenamefont {Ewart}
  \emph {et~al.}}]{ewart2025relaxation}%
  \BibitemOpen
  \bibfield  {author} {\bibinfo {author} {\bibfnamefont {R.~J.}\ \bibnamefont
  {Ewart}} \emph {et~al.},\ }\bibfield  {title} {\bibinfo {title} {Relaxation
  to universal non-maxwellian equilibria in a collisionless plasma},\
  }\href@noop {} {\bibfield  {journal} {\bibinfo  {journal} {Proceedings of the
  National Academy of Sciences}\ }\textbf {\bibinfo {volume} {122}},\ \bibinfo
  {pages} {e2417813122} (\bibinfo {year} {2025})}\BibitemShut {NoStop}%
\bibitem [{\citenamefont {Zhu}\ \emph {et~al.}(2025)\citenamefont {Zhu} \emph
  {et~al.}}]{zhu2025polarization}%
  \BibitemOpen
  \bibfield  {author} {\bibinfo {author} {\bibfnamefont {X.}~\bibnamefont
  {Zhu}} \emph {et~al.},\ }\bibfield  {title} {\bibinfo {title} {The
  polarization signatures of the inverse cascade in magnetic turbulence},\
  }\href@noop {} {\bibfield  {journal} {\bibinfo  {journal} {The Astrophysical
  Journal}\ }\textbf {\bibinfo {volume} {981}},\ \bibinfo {pages} {59}
  (\bibinfo {year} {2025})}\BibitemShut {NoStop}%
\bibitem [{\citenamefont {Gorbunov}\ \emph {et~al.}(2025)\citenamefont
  {Gorbunov}, \citenamefont {Gro{\v{s}}elj},\ and\ \citenamefont
  {Bacchini}}]{gorbunov2025leaking}%
  \BibitemOpen
  \bibfield  {author} {\bibinfo {author} {\bibfnamefont {E.~A.}\ \bibnamefont
  {Gorbunov}}, \bibinfo {author} {\bibfnamefont {D.}~\bibnamefont
  {Gro{\v{s}}elj}},\ and\ \bibinfo {author} {\bibfnamefont {F.}~\bibnamefont
  {Bacchini}},\ }\bibfield  {title} {\bibinfo {title} {Leaking outside the box:
  kinetic turbulence with cosmic-ray escape},\ }\href@noop {} {\bibfield
  {journal} {\bibinfo  {journal} {Physical Review Letters}\ }\textbf {\bibinfo
  {volume} {135}},\ \bibinfo {pages} {065201} (\bibinfo {year}
  {2025})}\BibitemShut {NoStop}%
\bibitem [{\citenamefont {Jenko}\ \emph {et~al.}(2000)\citenamefont {Jenko}
  \emph {et~al.}}]{Jenko2000}%
  \BibitemOpen
  \bibfield  {author} {\bibinfo {author} {\bibfnamefont {F.}~\bibnamefont
  {Jenko}} \emph {et~al.},\ }\bibfield  {title} {\bibinfo {title} {Electron
  temperature gradient driven turbulence},\ }\href@noop {} {\bibfield
  {journal} {\bibinfo  {journal} {Physics of Plasmas}\ }\textbf {\bibinfo
  {volume} {7}},\ \bibinfo {pages} {1904} (\bibinfo {year} {2000})}\BibitemShut
  {NoStop}%
\bibitem [{\citenamefont {Mattoo}\ \emph {et~al.}(2012)\citenamefont {Mattoo}
  \emph {et~al.}}]{Mattoo2012ETG}%
  \BibitemOpen
  \bibfield  {author} {\bibinfo {author} {\bibfnamefont {S.~K.}\ \bibnamefont
  {Mattoo}} \emph {et~al.},\ }\bibfield  {title} {\bibinfo {title}
  {Experimental observation of electron-temperature-gradient turbulence in a
  laboratory plasma},\ }\href@noop {} {\bibfield  {journal} {\bibinfo
  {journal} {Physical Review Letters}\ }\textbf {\bibinfo {volume} {108}},\
  \bibinfo {pages} {255007} (\bibinfo {year} {2012})}\BibitemShut {NoStop}%
\bibitem [{\citenamefont {Ongena}\ \emph {et~al.}(2016)\citenamefont {Ongena}
  \emph {et~al.}}]{ongena2016magnetic}%
  \BibitemOpen
  \bibfield  {author} {\bibinfo {author} {\bibfnamefont {J.}~\bibnamefont
  {Ongena}} \emph {et~al.},\ }\bibfield  {title} {\bibinfo {title}
  {Magnetic-confinement fusion},\ }\href@noop {} {\bibfield  {journal}
  {\bibinfo  {journal} {Nature Physics}\ }\textbf {\bibinfo {volume} {12}},\
  \bibinfo {pages} {398} (\bibinfo {year} {2016})}\BibitemShut {NoStop}%
\bibitem [{\citenamefont {Fiuza}\ \emph {et~al.}(2020)\citenamefont {Fiuza},
  \citenamefont {Swadling}, \citenamefont {Grassi} \emph {et~al.}}]{Fiuza2020}%
  \BibitemOpen
  \bibfield  {author} {\bibinfo {author} {\bibfnamefont {F.}~\bibnamefont
  {Fiuza}}, \bibinfo {author} {\bibfnamefont {G.~F.}\ \bibnamefont {Swadling}},
  \bibinfo {author} {\bibfnamefont {A.}~\bibnamefont {Grassi}}, \emph
  {et~al.},\ }\bibfield  {title} {\bibinfo {title} {Electron acceleration in
  laboratory-produced turbulent collisionless shocks},\ }\href@noop {}
  {\bibfield  {journal} {\bibinfo  {journal} {Nature Physics}\ }\textbf
  {\bibinfo {volume} {16}},\ \bibinfo {pages} {916} (\bibinfo {year}
  {2020})}\BibitemShut {NoStop}%
\bibitem [{\citenamefont {Yuan}\ \emph {et~al.}(2024)\citenamefont {Yuan} \emph
  {et~al.}}]{Yuan2024}%
  \BibitemOpen
  \bibfield  {author} {\bibinfo {author} {\bibfnamefont {D.}~\bibnamefont
  {Yuan}} \emph {et~al.},\ }\bibfield  {title} {\bibinfo {title} {Electron
  stochastic acceleration in laboratory-produced kinetic turbulent plasmas},\
  }\href@noop {} {\bibfield  {journal} {\bibinfo  {journal} {Nature
  Communications}\ }\textbf {\bibinfo {volume} {15}},\ \bibinfo {pages} {5897}
  (\bibinfo {year} {2024})}\BibitemShut {NoStop}%
\bibitem [{\citenamefont {Liu}\ \emph {et~al.}(2024)\citenamefont {Liu} \emph
  {et~al.}}]{Liu2024WeibelTurb}%
  \BibitemOpen
  \bibfield  {author} {\bibinfo {author} {\bibfnamefont {P.}~\bibnamefont
  {Liu}} \emph {et~al.},\ }\bibfield  {title} {\bibinfo {title} {Ion kinetics
  and neutron generation associated with electromagnetic turbulence in
  laboratory-scale counterstreaming plasmas},\ }\href@noop {} {\bibfield
  {journal} {\bibinfo  {journal} {Phys. Rev. Lett.}\ }\textbf {\bibinfo
  {volume} {132}},\ \bibinfo {pages} {155103} (\bibinfo {year}
  {2024})}\BibitemShut {NoStop}%
\bibitem [{\citenamefont {Bruno}\ and\ \citenamefont
  {Carbone}(2016)}]{Roberto2016solar}%
  \BibitemOpen
  \bibfield  {author} {\bibinfo {author} {\bibfnamefont {R.}~\bibnamefont
  {Bruno}}\ and\ \bibinfo {author} {\bibfnamefont {V.}~\bibnamefont
  {Carbone}},\ }\href@noop {} {\emph {\bibinfo {title} {Turbulence in the Solar
  Wind}}}\ (\bibinfo  {publisher} {Springer},\ \bibinfo {year}
  {2016})\BibitemShut {NoStop}%
\bibitem [{\citenamefont {Kawazura}\ \emph {et~al.}(2020)\citenamefont
  {Kawazura}, \citenamefont {Schekochihin}, \citenamefont {Barnes},
  \citenamefont {TenBarge}, \citenamefont {Tong}, \citenamefont {Klein},\ and\
  \citenamefont {Dorland}}]{kawazura2020ion}%
  \BibitemOpen
  \bibfield  {author} {\bibinfo {author} {\bibfnamefont {Y.}~\bibnamefont
  {Kawazura}}, \bibinfo {author} {\bibfnamefont {A.}~\bibnamefont
  {Schekochihin}}, \bibinfo {author} {\bibfnamefont {M.}~\bibnamefont
  {Barnes}}, \bibinfo {author} {\bibfnamefont {J.}~\bibnamefont {TenBarge}},
  \bibinfo {author} {\bibfnamefont {Y.}~\bibnamefont {Tong}}, \bibinfo {author}
  {\bibfnamefont {K.}~\bibnamefont {Klein}},\ and\ \bibinfo {author}
  {\bibfnamefont {W.}~\bibnamefont {Dorland}},\ }\bibfield  {title} {\bibinfo
  {title} {Ion versus electron heating in compressively driven astrophysical
  gyrokinetic turbulence},\ }\href@noop {} {\bibfield  {journal} {\bibinfo
  {journal} {Physical Review X}\ }\textbf {\bibinfo {volume} {10}},\ \bibinfo
  {pages} {041050} (\bibinfo {year} {2020})}\BibitemShut {NoStop}%
\bibitem [{\citenamefont {Bowen}\ \emph {et~al.}(2025)\citenamefont {Bowen},
  \citenamefont {Ervin}, \citenamefont {Mallet} \emph
  {et~al.}}]{Bowen2025solar}%
  \BibitemOpen
  \bibfield  {author} {\bibinfo {author} {\bibfnamefont {T.~A.}\ \bibnamefont
  {Bowen}}, \bibinfo {author} {\bibfnamefont {T.}~\bibnamefont {Ervin}},
  \bibinfo {author} {\bibfnamefont {A.}~\bibnamefont {Mallet}}, \emph
  {et~al.},\ }\bibfield  {title} {\bibinfo {title} {Stochastic heating in the
  sub-alfv\'enic solar wind},\ }\href@noop {} {\bibfield  {journal} {\bibinfo
  {journal} {Physical review letters}\ }\textbf {\bibinfo {volume} {135}},\
  \bibinfo {pages} {255201} (\bibinfo {year} {2025})}\BibitemShut {NoStop}%
\bibitem [{\citenamefont {Marscher}\ \emph {et~al.}(2008)\citenamefont
  {Marscher} \emph {et~al.}}]{Marscher2008AGN}%
  \BibitemOpen
  \bibfield  {author} {\bibinfo {author} {\bibfnamefont {A.~P.}\ \bibnamefont
  {Marscher}} \emph {et~al.},\ }\bibfield  {title} {\bibinfo {title} {The inner
  jet of an active galactic nucleus as revealed by a radio-to-$\gamma$-ray
  outburs},\ }\href@noop {} {\bibfield  {journal} {\bibinfo  {journal}
  {Nature}\ }\textbf {\bibinfo {volume} {452}},\ \bibinfo {pages} {966}
  (\bibinfo {year} {2008})}\BibitemShut {NoStop}%
\bibitem [{\citenamefont {Madejski}\ and\ \citenamefont
  {Sikora}(2016)}]{Madejski2016}%
  \BibitemOpen
  \bibfield  {author} {\bibinfo {author} {\bibfnamefont {G.~G.}\ \bibnamefont
  {Madejski}}\ and\ \bibinfo {author} {\bibfnamefont {M.}~\bibnamefont
  {Sikora}},\ }\bibfield  {title} {\bibinfo {title} {Gamma-ray observations of
  active galactic nuclei},\ }\href@noop {} {\bibfield  {journal} {\bibinfo
  {journal} {Annual Review of Astronomy and Astrophysics}\ }\textbf {\bibinfo
  {volume} {54}},\ \bibinfo {pages} {725} (\bibinfo {year} {2016})}\BibitemShut
  {NoStop}%
\bibitem [{\citenamefont {Bourne}\ and\ \citenamefont
  {Sijacki}(2017)}]{Bourne2017}%
  \BibitemOpen
  \bibfield  {author} {\bibinfo {author} {\bibfnamefont {M.~A.}\ \bibnamefont
  {Bourne}}\ and\ \bibinfo {author} {\bibfnamefont {D.}~\bibnamefont
  {Sijacki}},\ }\bibfield  {title} {\bibinfo {title} {Agn jet feedback on a
  moving mesh: cocoon inflation, gas flows and turbulence},\ }\href@noop {}
  {\bibfield  {journal} {\bibinfo  {journal} {Monthly Notices of the Royal
  Astronomical Society}\ }\textbf {\bibinfo {volume} {472}},\ \bibinfo {pages}
  {4707} (\bibinfo {year} {2017})}\BibitemShut {NoStop}%
\bibitem [{\citenamefont {Piran}(2005)}]{gamma_ray_burst}%
  \BibitemOpen
  \bibfield  {author} {\bibinfo {author} {\bibfnamefont {T.}~\bibnamefont
  {Piran}},\ }\bibfield  {title} {\bibinfo {title} {The physics of gamma-ray
  bursts},\ }\href@noop {} {\bibfield  {journal} {\bibinfo  {journal} {Reviews
  of Modern Physics}\ }\textbf {\bibinfo {volume} {76}},\ \bibinfo {pages}
  {1143} (\bibinfo {year} {2005})}\BibitemShut {NoStop}%
\bibitem [{\citenamefont {Mészáros}(2006)}]{Meszaros2006}%
  \BibitemOpen
  \bibfield  {author} {\bibinfo {author} {\bibfnamefont {P.}~\bibnamefont
  {Mészáros}},\ }\bibfield  {title} {\bibinfo {title} {Gamma-ray bursts},\
  }\href@noop {} {\bibfield  {journal} {\bibinfo  {journal} {Rep. Prog. Phys.}\
  }\textbf {\bibinfo {volume} {69}},\ \bibinfo {pages} {2259} (\bibinfo {year}
  {2006})}\BibitemShut {NoStop}%
\bibitem [{\citenamefont {Burgess}\ \emph {et~al.}(2020)\citenamefont {Burgess}
  \emph {et~al.}}]{Burgess2020}%
  \BibitemOpen
  \bibfield  {author} {\bibinfo {author} {\bibfnamefont {J.~M.}\ \bibnamefont
  {Burgess}} \emph {et~al.},\ }\bibfield  {title} {\bibinfo {title} {Gamma-ray
  bursts as cool synchrotron sources},\ }\href@noop {} {\bibfield  {journal}
  {\bibinfo  {journal} {Nature Astronomy}\ }\textbf {\bibinfo {volume} {4}},\
  \bibinfo {pages} {174} (\bibinfo {year} {2020})}\BibitemShut {NoStop}%
\bibitem [{\citenamefont {Aab}\ \emph {et~al.}(2020)\citenamefont {Aab} \emph
  {et~al.}}]{Aab2020Observatory}%
  \BibitemOpen
  \bibfield  {author} {\bibinfo {author} {\bibfnamefont {A.}~\bibnamefont
  {Aab}} \emph {et~al.} (\bibinfo {collaboration} {The Pierre Auger
  Collaboration}),\ }\bibfield  {title} {\bibinfo {title} {Features of the
  energy spectrum of cosmic rays above
  $2.5\ifmmode\times\else\texttimes\fi{}{10}^{18}\text{ }\text{ }\mathrm{eV}$
  using the pierre auger observatory},\ }\href@noop {} {\bibfield  {journal}
  {\bibinfo  {journal} {Physical Review Letters}\ }\textbf {\bibinfo {volume}
  {125}},\ \bibinfo {pages} {121106} (\bibinfo {year} {2020})}\BibitemShut
  {NoStop}%
\bibitem [{\citenamefont {Abdul~Halim}\ \emph {et~al.}(2023)\citenamefont
  {Abdul~Halim} \emph {et~al.}}]{Abdul2023Observatory}%
  \BibitemOpen
  \bibfield  {author} {\bibinfo {author} {\bibfnamefont {A.}~\bibnamefont
  {Abdul~Halim}} \emph {et~al.},\ }\bibfield  {title} {\bibinfo {title}
  {Constraining the sources of ultra-high-energy cosmic rays across and above
  the ankle with the spectrum and composition data measured at the pierre auger
  observatory},\ }\href@noop {} {\bibfield  {journal} {\bibinfo  {journal}
  {Journal of Cosmology and Astroparticle Physics}\ }\textbf {\bibinfo {volume}
  {2023}}\bibinfo  {number} { (05)},\ \bibinfo {pages} {024}}\BibitemShut
  {NoStop}%
\bibitem [{\citenamefont {Michel}(1982)}]{michel1982theory}%
  \BibitemOpen
\bibfield  {number} {  }\bibfield  {author} {\bibinfo {author} {\bibfnamefont
  {F.~C.}\ \bibnamefont {Michel}},\ }\bibfield  {title} {\bibinfo {title}
  {Theory of pulsar magnetospheres},\ }\href@noop {} {\bibfield  {journal}
  {\bibinfo  {journal} {Reviews of Modern Physics}\ }\textbf {\bibinfo {volume}
  {54}},\ \bibinfo {pages} {1} (\bibinfo {year} {1982})}\BibitemShut {NoStop}%
\bibitem [{\citenamefont {Uzdensky}\ and\ \citenamefont
  {Spitkovsky}(2014)}]{uzdensky2014physical}%
  \BibitemOpen
  \bibfield  {author} {\bibinfo {author} {\bibfnamefont {D.~A.}\ \bibnamefont
  {Uzdensky}}\ and\ \bibinfo {author} {\bibfnamefont {A.}~\bibnamefont
  {Spitkovsky}},\ }\bibfield  {title} {\bibinfo {title} {Physical conditions in
  the reconnection layer in pulsar magnetospheres},\ }\href@noop {} {\bibfield
  {journal} {\bibinfo  {journal} {The Astrophysical Journal}\ }\textbf
  {\bibinfo {volume} {780}},\ \bibinfo {pages} {3} (\bibinfo {year}
  {2014})}\BibitemShut {NoStop}%
\bibitem [{\citenamefont {Philippov}\ and\ \citenamefont
  {Kramer}(2022)}]{philippov2022pulsar}%
  \BibitemOpen
  \bibfield  {author} {\bibinfo {author} {\bibfnamefont {A.}~\bibnamefont
  {Philippov}}\ and\ \bibinfo {author} {\bibfnamefont {M.}~\bibnamefont
  {Kramer}},\ }\bibfield  {title} {\bibinfo {title} {Pulsar magnetospheres and
  their radiation},\ }\href@noop {} {\bibfield  {journal} {\bibinfo  {journal}
  {Annual Review of Astronomy and Astrophysics}\ }\textbf {\bibinfo {volume}
  {60}},\ \bibinfo {pages} {495} (\bibinfo {year} {2022})}\BibitemShut
  {NoStop}%
\bibitem [{\citenamefont {Kaspi}\ and\ \citenamefont
  {Beloborodov}(2017)}]{kaspi2017magnetars}%
  \BibitemOpen
  \bibfield  {author} {\bibinfo {author} {\bibfnamefont {V.~M.}\ \bibnamefont
  {Kaspi}}\ and\ \bibinfo {author} {\bibfnamefont {A.~M.}\ \bibnamefont
  {Beloborodov}},\ }\bibfield  {title} {\bibinfo {title} {Magnetars},\
  }\href@noop {} {\bibfield  {journal} {\bibinfo  {journal} {Annual Review of
  Astronomy and Astrophysics}\ }\textbf {\bibinfo {volume} {55}},\ \bibinfo
  {pages} {261} (\bibinfo {year} {2017})}\BibitemShut {NoStop}%
\bibitem [{\citenamefont {Parfrey}\ \emph {et~al.}(2013)\citenamefont
  {Parfrey}, \citenamefont {Beloborodov},\ and\ \citenamefont
  {Hui}}]{Parfrey2013MS}%
  \BibitemOpen
  \bibfield  {author} {\bibinfo {author} {\bibfnamefont {K.}~\bibnamefont
  {Parfrey}}, \bibinfo {author} {\bibfnamefont {A.~M.}\ \bibnamefont
  {Beloborodov}},\ and\ \bibinfo {author} {\bibfnamefont {L.}~\bibnamefont
  {Hui}},\ }\bibfield  {title} {\bibinfo {title} {Dynamics of strongly twisted
  relativistic magnetospheres},\ }\href@noop {} {\bibfield  {journal} {\bibinfo
   {journal} {The Astrophysical Journal}\ }\textbf {\bibinfo {volume} {774}},\
  \bibinfo {pages} {92} (\bibinfo {year} {2013})}\BibitemShut {NoStop}%
\bibitem [{\citenamefont {Li}\ \emph {et~al.}(2016{\natexlab{a}})\citenamefont
  {Li}, \citenamefont {Levin},\ and\ \citenamefont {Beloborodov}}]{Li2016MS}%
  \BibitemOpen
  \bibfield  {author} {\bibinfo {author} {\bibfnamefont {X.}~\bibnamefont
  {Li}}, \bibinfo {author} {\bibfnamefont {Y.}~\bibnamefont {Levin}},\ and\
  \bibinfo {author} {\bibfnamefont {A.~M.}\ \bibnamefont {Beloborodov}},\
  }\bibfield  {title} {\bibinfo {title} {Magnetar outbursts from avalanches of
  hall waves and crustal failures},\ }\href@noop {} {\bibfield  {journal}
  {\bibinfo  {journal} {The Astrophysical Journal}\ }\textbf {\bibinfo {volume}
  {833}},\ \bibinfo {pages} {189} (\bibinfo {year}
  {2016}{\natexlab{a}})}\BibitemShut {NoStop}%
\bibitem [{\citenamefont {Schoeffler}\ \emph {et~al.}(2019)\citenamefont
  {Schoeffler}, \citenamefont {Grismayer}, \citenamefont {Uzdensky},
  \citenamefont {Fonseca},\ and\ \citenamefont {Silva}}]{Schoeffler2019MS}%
  \BibitemOpen
  \bibfield  {author} {\bibinfo {author} {\bibfnamefont {K.~M.}\ \bibnamefont
  {Schoeffler}}, \bibinfo {author} {\bibfnamefont {T.}~\bibnamefont
  {Grismayer}}, \bibinfo {author} {\bibfnamefont {D.}~\bibnamefont {Uzdensky}},
  \bibinfo {author} {\bibfnamefont {R.~A.}\ \bibnamefont {Fonseca}},\ and\
  \bibinfo {author} {\bibfnamefont {L.~O.}\ \bibnamefont {Silva}},\ }\bibfield
  {title} {\bibinfo {title} {Bright gamma-ray flares powered by magnetic
  reconnection in qed-strength magnetic fields},\ }\href@noop {} {\bibfield
  {journal} {\bibinfo  {journal} {The Astrophysical Journal}\ }\textbf
  {\bibinfo {volume} {870}},\ \bibinfo {pages} {49} (\bibinfo {year}
  {2019})}\BibitemShut {NoStop}%
\bibitem [{\citenamefont {Thompson}\ and\ \citenamefont
  {Duncan}(1995)}]{Thompson1995MNRAS}%
  \BibitemOpen
  \bibfield  {author} {\bibinfo {author} {\bibfnamefont {C.}~\bibnamefont
  {Thompson}}\ and\ \bibinfo {author} {\bibfnamefont {R.~C.}\ \bibnamefont
  {Duncan}},\ }\bibfield  {title} {\bibinfo {title} {The soft gamma repeaters
  as very strongly magnetized neutron stars - i. radiative mechanism for
  outbursts},\ }\href@noop {} {\bibfield  {journal} {\bibinfo  {journal}
  {Monthly Notices of the Royal Astronomical Society}\ }\textbf {\bibinfo
  {volume} {275}},\ \bibinfo {pages} {255} (\bibinfo {year}
  {1995})}\BibitemShut {NoStop}%
\bibitem [{\citenamefont {Yang}\ and\ \citenamefont
  {Zhang}(2015)}]{Yang2015Polarization}%
  \BibitemOpen
  \bibfield  {author} {\bibinfo {author} {\bibfnamefont {Y.-P.}\ \bibnamefont
  {Yang}}\ and\ \bibinfo {author} {\bibfnamefont {B.}~\bibnamefont {Zhang}},\
  }\bibfield  {title} {\bibinfo {title} {On the polarization properties of
  magnetar giant flare pulsating tails},\ }\href@noop {} {\bibfield  {journal}
  {\bibinfo  {journal} {The Astrophysical Journal}\ }\textbf {\bibinfo {volume}
  {815}},\ \bibinfo {pages} {45} (\bibinfo {year} {2015})}\BibitemShut
  {NoStop}%
\bibitem [{\citenamefont {Blaes}\ \emph {et~al.}(1989)\citenamefont {Blaes},
  \citenamefont {Blandford}, \citenamefont {Goldreich},\ and\ \citenamefont
  {Madau}}]{blaes1989neutron_starvation}%
  \BibitemOpen
  \bibfield  {author} {\bibinfo {author} {\bibfnamefont {O.}~\bibnamefont
  {Blaes}}, \bibinfo {author} {\bibfnamefont {R.}~\bibnamefont {Blandford}},
  \bibinfo {author} {\bibfnamefont {P.}~\bibnamefont {Goldreich}},\ and\
  \bibinfo {author} {\bibfnamefont {P.}~\bibnamefont {Madau}},\ }\bibfield
  {title} {\bibinfo {title} {Neutron starquake models for gamma-ray bursts},\
  }\href@noop {} {\bibfield  {journal} {\bibinfo  {journal} {Astrophysical
  Journal, Part 1 (ISSN 0004-637X), vol. 343, Aug. 15, 1989, p. 839-848.}\
  }\textbf {\bibinfo {volume} {343}},\ \bibinfo {pages} {839} (\bibinfo {year}
  {1989})}\BibitemShut {NoStop}%
\bibitem [{\citenamefont {Li}\ \emph {et~al.}(2016{\natexlab{b}})\citenamefont
  {Li}, \citenamefont {Levin},\ and\ \citenamefont
  {Beloborodov}}]{LiMagnetar2016}%
  \BibitemOpen
  \bibfield  {author} {\bibinfo {author} {\bibfnamefont {X.}~\bibnamefont
  {Li}}, \bibinfo {author} {\bibfnamefont {Y.}~\bibnamefont {Levin}},\ and\
  \bibinfo {author} {\bibfnamefont {A.~M.}\ \bibnamefont {Beloborodov}},\
  }\bibfield  {title} {\bibinfo {title} {Magnetar outbursts from avalanches of
  hall waves and crustal failures},\ }\href@noop {} {\bibfield  {journal}
  {\bibinfo  {journal} {The Astrophysical Journal}\ }\textbf {\bibinfo {volume}
  {833}},\ \bibinfo {pages} {189} (\bibinfo {year}
  {2016}{\natexlab{b}})}\BibitemShut {NoStop}%
\bibitem [{\citenamefont {Yuan}\ \emph {et~al.}(2020)\citenamefont {Yuan},
  \citenamefont {Beloborodov}, \citenamefont {Chen},\ and\ \citenamefont
  {Levin}}]{YuanMagnetarQuake2020}%
  \BibitemOpen
  \bibfield  {author} {\bibinfo {author} {\bibfnamefont {Y.}~\bibnamefont
  {Yuan}}, \bibinfo {author} {\bibfnamefont {A.~M.}\ \bibnamefont
  {Beloborodov}}, \bibinfo {author} {\bibfnamefont {A.~Y.}\ \bibnamefont
  {Chen}},\ and\ \bibinfo {author} {\bibfnamefont {Y.}~\bibnamefont {Levin}},\
  }\bibfield  {title} {\bibinfo {title} {Plasmoid ejection by alfvén waves and
  the fast radio bursts from sgr 1935+2154},\ }\href@noop {} {\bibfield
  {journal} {\bibinfo  {journal} {The Astrophysical Journal Letters}\ }\textbf
  {\bibinfo {volume} {900}},\ \bibinfo {pages} {L21} (\bibinfo {year}
  {2020})}\BibitemShut {NoStop}%
\bibitem [{\citenamefont {Bransgrove}\ \emph {et~al.}(2020)\citenamefont
  {Bransgrove}, \citenamefont {Beloborodov},\ and\ \citenamefont
  {Levin}}]{Bransgrove2020QuakeAlfven}%
  \BibitemOpen
  \bibfield  {author} {\bibinfo {author} {\bibfnamefont {A.}~\bibnamefont
  {Bransgrove}}, \bibinfo {author} {\bibfnamefont {A.~M.}\ \bibnamefont
  {Beloborodov}},\ and\ \bibinfo {author} {\bibfnamefont {Y.}~\bibnamefont
  {Levin}},\ }\bibfield  {title} {\bibinfo {title} {A quake quenching the vela
  pulsar},\ }\href@noop {} {\bibfield  {journal} {\bibinfo  {journal} {The
  Astrophysical Journal}\ }\textbf {\bibinfo {volume} {897}},\ \bibinfo {pages}
  {173} (\bibinfo {year} {2020})}\BibitemShut {NoStop}%
\bibitem [{\citenamefont {Chen}\ \emph
  {et~al.}(2022{\natexlab{a}})\citenamefont {Chen} \emph
  {et~al.}}]{ChenStarvation2022}%
  \BibitemOpen
  \bibfield  {author} {\bibinfo {author} {\bibfnamefont {A.~Y.}\ \bibnamefont
  {Chen}} \emph {et~al.},\ }\bibfield  {title} {\bibinfo {title} {Relativistic
  alfvén waves entering charge-starvation in the magnetospheres of neutron
  stars},\ }\href@noop {} {\bibfield  {journal} {\bibinfo  {journal} {The
  Astrophysical Journal}\ }\textbf {\bibinfo {volume} {929}},\ \bibinfo {pages}
  {31} (\bibinfo {year} {2022}{\natexlab{a}})}\BibitemShut {NoStop}%
\bibitem [{\citenamefont {Thompson}\ and\ \citenamefont
  {Blaes}(1998)}]{ThompsonStarvation1998}%
  \BibitemOpen
  \bibfield  {author} {\bibinfo {author} {\bibfnamefont {C.}~\bibnamefont
  {Thompson}}\ and\ \bibinfo {author} {\bibfnamefont {O.}~\bibnamefont
  {Blaes}},\ }\bibfield  {title} {\bibinfo {title} {Magnetohydrodynamics in the
  extreme relativistic limit},\ }\href@noop {} {\bibfield  {journal} {\bibinfo
  {journal} {Phys. Rev. D}\ }\textbf {\bibinfo {volume} {57}},\ \bibinfo
  {pages} {3219} (\bibinfo {year} {1998})}\BibitemShut {NoStop}%
\bibitem [{\citenamefont {Li}\ \emph {et~al.}(2019{\natexlab{a}})\citenamefont
  {Li}, \citenamefont {Zrake},\ and\ \citenamefont
  {Beloborodov}}]{Li2019AlfvenTurb}%
  \BibitemOpen
  \bibfield  {author} {\bibinfo {author} {\bibfnamefont {X.}~\bibnamefont
  {Li}}, \bibinfo {author} {\bibfnamefont {J.}~\bibnamefont {Zrake}},\ and\
  \bibinfo {author} {\bibfnamefont {A.~M.}\ \bibnamefont {Beloborodov}},\
  }\bibfield  {title} {\bibinfo {title} {Dissipation of alfv\'{e}n waves in
  relativistic magnetospheres of magnetars},\ }\href@noop {} {\bibfield
  {journal} {\bibinfo  {journal} {The Astrophysical Journal}\ }\textbf
  {\bibinfo {volume} {881}},\ \bibinfo {pages} {13} (\bibinfo {year}
  {2019}{\natexlab{a}})}\BibitemShut {NoStop}%
\bibitem [{\citenamefont {Zhdankin}\ \emph {et~al.}(2020)\citenamefont
  {Zhdankin}, \citenamefont {Uzdensky}, \citenamefont {Werner},\ and\
  \citenamefont {Begelman}}]{zhdankin2020kinetic}%
  \BibitemOpen
  \bibfield  {author} {\bibinfo {author} {\bibfnamefont {V.}~\bibnamefont
  {Zhdankin}}, \bibinfo {author} {\bibfnamefont {D.~A.}\ \bibnamefont
  {Uzdensky}}, \bibinfo {author} {\bibfnamefont {G.~R.}\ \bibnamefont
  {Werner}},\ and\ \bibinfo {author} {\bibfnamefont {M.~C.}\ \bibnamefont
  {Begelman}},\ }\bibfield  {title} {\bibinfo {title} {Kinetic turbulence in
  shining pair plasma: intermittent beaming and thermalization by radiative
  cooling},\ }\href@noop {} {\bibfield  {journal} {\bibinfo  {journal} {Monthly
  Notices of the Royal Astronomical Society}\ }\textbf {\bibinfo {volume}
  {493}},\ \bibinfo {pages} {603} (\bibinfo {year} {2020})}\BibitemShut
  {NoStop}%
\bibitem [{\citenamefont {Comisso}\ and\ \citenamefont
  {Sironi}(2021)}]{Comisso2021}%
  \BibitemOpen
  \bibfield  {author} {\bibinfo {author} {\bibfnamefont {L.}~\bibnamefont
  {Comisso}}\ and\ \bibinfo {author} {\bibfnamefont {L.}~\bibnamefont
  {Sironi}},\ }\bibfield  {title} {\bibinfo {title} {Pitch-angle anisotropy
  controls particle acceleration and cooling in radiative relativistic plasma
  turbulence},\ }\href@noop {} {\bibfield  {journal} {\bibinfo  {journal}
  {Physical Review Letters}\ }\textbf {\bibinfo {volume} {127}},\ \bibinfo
  {pages} {255102} (\bibinfo {year} {2021})}\BibitemShut {NoStop}%
\bibitem [{\citenamefont {N{\"a}ttil{\"a}}\ and\ \citenamefont
  {Beloborodov}(2021)}]{nattila2021radiative}%
  \BibitemOpen
  \bibfield  {author} {\bibinfo {author} {\bibfnamefont {J.}~\bibnamefont
  {N{\"a}ttil{\"a}}}\ and\ \bibinfo {author} {\bibfnamefont {A.~M.}\
  \bibnamefont {Beloborodov}},\ }\bibfield  {title} {\bibinfo {title}
  {Radiative turbulent flares in magnetically dominated plasmas},\ }\href@noop
  {} {\bibfield  {journal} {\bibinfo  {journal} {The Astrophysical Journal}\
  }\textbf {\bibinfo {volume} {921}},\ \bibinfo {pages} {87} (\bibinfo {year}
  {2021})}\BibitemShut {NoStop}%
\bibitem [{\citenamefont {Mehlhaff}\ \emph {et~al.}(2025)\citenamefont
  {Mehlhaff}, \citenamefont {Zhou},\ and\ \citenamefont
  {Zhdankin}}]{mehlhaff2025radiative}%
  \BibitemOpen
  \bibfield  {author} {\bibinfo {author} {\bibfnamefont {J.~M.}\ \bibnamefont
  {Mehlhaff}}, \bibinfo {author} {\bibfnamefont {M.}~\bibnamefont {Zhou}},\
  and\ \bibinfo {author} {\bibfnamefont {V.}~\bibnamefont {Zhdankin}},\
  }\bibfield  {title} {\bibinfo {title} {Radiative relativistic turbulence as
  an in situ pair-plasma source in blazar jets},\ }\href@noop {} {\bibfield
  {journal} {\bibinfo  {journal} {The Astrophysical Journal}\ }\textbf
  {\bibinfo {volume} {987}},\ \bibinfo {pages} {159} (\bibinfo {year}
  {2025})}\BibitemShut {NoStop}%
\bibitem [{\citenamefont {Bilbao}\ \emph {et~al.}(2025)\citenamefont {Bilbao},
  \citenamefont {Silva},\ and\ \citenamefont {Silva}}]{bilbao2025radiative}%
  \BibitemOpen
  \bibfield  {author} {\bibinfo {author} {\bibfnamefont {P.~J.}\ \bibnamefont
  {Bilbao}}, \bibinfo {author} {\bibfnamefont {T.}~\bibnamefont {Silva}},\ and\
  \bibinfo {author} {\bibfnamefont {L.~O.}\ \bibnamefont {Silva}},\ }\bibfield
  {title} {\bibinfo {title} {Radiative cooling induced coherent maser emission
  in relativistic plasmas},\ }\href@noop {} {\bibfield  {journal} {\bibinfo
  {journal} {Science advances}\ }\textbf {\bibinfo {volume} {11}},\ \bibinfo
  {pages} {8912} (\bibinfo {year} {2025})}\BibitemShut {NoStop}%
\bibitem [{\citenamefont {Di~Piazza}\ \emph {et~al.}(2012)\citenamefont
  {Di~Piazza}, \citenamefont {M{\"u}ller}, \citenamefont {Hatsagortsyan},\ and\
  \citenamefont {Keitel}}]{di2012extremely}%
  \BibitemOpen
  \bibfield  {author} {\bibinfo {author} {\bibfnamefont {A.}~\bibnamefont
  {Di~Piazza}}, \bibinfo {author} {\bibfnamefont {C.}~\bibnamefont
  {M{\"u}ller}}, \bibinfo {author} {\bibfnamefont {K.~Z.}\ \bibnamefont
  {Hatsagortsyan}},\ and\ \bibinfo {author} {\bibfnamefont {C.~H.}\
  \bibnamefont {Keitel}},\ }\bibfield  {title} {\bibinfo {title} {Extremely
  high-intensity laser interactions with fundamental quantum systems},\
  }\href@noop {} {\bibfield  {journal} {\bibinfo  {journal} {Reviews of Modern
  Physics}\ }\textbf {\bibinfo {volume} {84}},\ \bibinfo {pages} {1177}
  (\bibinfo {year} {2012})}\BibitemShut {NoStop}%
\bibitem [{\citenamefont {Uzdensky}\ and\ \citenamefont
  {Rightley}(2014)}]{uzdensky2014plasma}%
  \BibitemOpen
  \bibfield  {author} {\bibinfo {author} {\bibfnamefont {D.~A.}\ \bibnamefont
  {Uzdensky}}\ and\ \bibinfo {author} {\bibfnamefont {S.}~\bibnamefont
  {Rightley}},\ }\bibfield  {title} {\bibinfo {title} {Plasma physics of
  extreme astrophysical environments},\ }\href@noop {} {\bibfield  {journal}
  {\bibinfo  {journal} {Reports on Progress in Physics}\ }\textbf {\bibinfo
  {volume} {77}},\ \bibinfo {pages} {036902} (\bibinfo {year}
  {2014})}\BibitemShut {NoStop}%
\bibitem [{\citenamefont {Fedotov}\ \emph {et~al.}(2023)\citenamefont
  {Fedotov}, \citenamefont {Ilderton}, \citenamefont {Karbstein}, \citenamefont
  {King}, \citenamefont {Seipt}, \citenamefont {Taya},\ and\ \citenamefont
  {Torgrimsson}}]{fedotov2023advances}%
  \BibitemOpen
  \bibfield  {author} {\bibinfo {author} {\bibfnamefont {A.}~\bibnamefont
  {Fedotov}}, \bibinfo {author} {\bibfnamefont {A.}~\bibnamefont {Ilderton}},
  \bibinfo {author} {\bibfnamefont {F.}~\bibnamefont {Karbstein}}, \bibinfo
  {author} {\bibfnamefont {B.}~\bibnamefont {King}}, \bibinfo {author}
  {\bibfnamefont {D.}~\bibnamefont {Seipt}}, \bibinfo {author} {\bibfnamefont
  {H.}~\bibnamefont {Taya}},\ and\ \bibinfo {author} {\bibfnamefont
  {G.}~\bibnamefont {Torgrimsson}},\ }\bibfield  {title} {\bibinfo {title}
  {Advances in qed with intense background fields},\ }\href@noop {} {\bibfield
  {journal} {\bibinfo  {journal} {Physics Reports}\ }\textbf {\bibinfo {volume}
  {1010}},\ \bibinfo {pages} {1} (\bibinfo {year} {2023})}\BibitemShut
  {NoStop}%
\bibitem [{\citenamefont {Baier}\ and\ \citenamefont
  {Katkov}(1967)}]{Baier_1967}%
  \BibitemOpen
  \bibfield  {author} {\bibinfo {author} {\bibfnamefont {V.~N.}\ \bibnamefont
  {Baier}}\ and\ \bibinfo {author} {\bibfnamefont {V.~M.}\ \bibnamefont
  {Katkov}},\ }\bibfield  {title} {\bibinfo {title} {{Radiational polarization
  of electrons in inhomogeneous magnetic field}},\ }\href@noop {} {\bibfield
  {journal} {\bibinfo  {journal} {Phys. Lett. A}\ }\textbf {\bibinfo {volume}
  {24}},\ \bibinfo {pages} {327} (\bibinfo {year} {1967})}\BibitemShut
  {NoStop}%
\bibitem [{\citenamefont {Del~Sorbo}\ \emph {et~al.}(2017)\citenamefont
  {Del~Sorbo} \emph {et~al.}}]{Sorbo_2017}%
  \BibitemOpen
  \bibfield  {author} {\bibinfo {author} {\bibfnamefont {D.}~\bibnamefont
  {Del~Sorbo}} \emph {et~al.},\ }\bibfield  {title} {\bibinfo {title} {Spin
  polarization of electrons by ultraintense lasers},\ }\href@noop {} {\bibfield
   {journal} {\bibinfo  {journal} {Phys. Rev. A}\ }\textbf {\bibinfo {volume}
  {96}},\ \bibinfo {pages} {043407} (\bibinfo {year} {2017})}\BibitemShut
  {NoStop}%
\bibitem [{\citenamefont {Seipt}\ \emph {et~al.}(2018)\citenamefont {Seipt},
  \citenamefont {Del~Sorbo}, \citenamefont {Ridgers},\ and\ \citenamefont
  {Thomas}}]{Seipt_2018}%
  \BibitemOpen
  \bibfield  {author} {\bibinfo {author} {\bibfnamefont {D.}~\bibnamefont
  {Seipt}}, \bibinfo {author} {\bibfnamefont {D.}~\bibnamefont {Del~Sorbo}},
  \bibinfo {author} {\bibfnamefont {C.~P.}\ \bibnamefont {Ridgers}},\ and\
  \bibinfo {author} {\bibfnamefont {A.~G.~R.}\ \bibnamefont {Thomas}},\
  }\bibfield  {title} {\bibinfo {title} {Theory of radiative electron
  polarization in strong laser fields},\ }\href@noop {} {\bibfield  {journal}
  {\bibinfo  {journal} {Phys. Rev. A}\ }\textbf {\bibinfo {volume} {98}},\
  \bibinfo {pages} {023417} (\bibinfo {year} {2018})}\BibitemShut {NoStop}%
\bibitem [{\citenamefont {Li}\ \emph {et~al.}(2019{\natexlab{b}})\citenamefont
  {Li} \emph {et~al.}}]{li2019ultrarelativistic}%
  \BibitemOpen
  \bibfield  {author} {\bibinfo {author} {\bibfnamefont {Y.-F.}\ \bibnamefont
  {Li}} \emph {et~al.},\ }\bibfield  {title} {\bibinfo {title}
  {Ultrarelativistic electron-beam polarization in single-shot interaction with
  an ultraintense laser pulse},\ }\href@noop {} {\bibfield  {journal} {\bibinfo
   {journal} {Physical review letters}\ }\textbf {\bibinfo {volume} {122}},\
  \bibinfo {pages} {154801} (\bibinfo {year} {2019}{\natexlab{b}})}\BibitemShut
  {NoStop}%
\bibitem [{\citenamefont {Geng}\ \emph {et~al.}(2020)\citenamefont {Geng},
  \citenamefont {Ji}, \citenamefont {Shen} \emph {et~al.}}]{geng2020spin}%
  \BibitemOpen
  \bibfield  {author} {\bibinfo {author} {\bibfnamefont {X.}~\bibnamefont
  {Geng}}, \bibinfo {author} {\bibfnamefont {L.}~\bibnamefont {Ji}}, \bibinfo
  {author} {\bibfnamefont {B.}~\bibnamefont {Shen}}, \emph {et~al.},\
  }\bibfield  {title} {\bibinfo {title} {Spin-dependent radiative deflection in
  the quantum radiation-reaction regime},\ }\href@noop {} {\bibfield  {journal}
  {\bibinfo  {journal} {New Journal of Physics}\ }\textbf {\bibinfo {volume}
  {22}},\ \bibinfo {pages} {013007} (\bibinfo {year} {2020})}\BibitemShut
  {NoStop}%
\bibitem [{\citenamefont {Li}\ \emph {et~al.}(2020{\natexlab{a}})\citenamefont
  {Li} \emph {et~al.}}]{li2020polarized}%
  \BibitemOpen
  \bibfield  {author} {\bibinfo {author} {\bibfnamefont {Y.-F.}\ \bibnamefont
  {Li}} \emph {et~al.},\ }\bibfield  {title} {\bibinfo {title} {Polarized
  ultrashort brilliant multi-gev $\gamma$ rays via single-shot laser-electron
  interaction},\ }\href@noop {} {\bibfield  {journal} {\bibinfo  {journal}
  {Physical Review Letters}\ }\textbf {\bibinfo {volume} {124}},\ \bibinfo
  {pages} {014801} (\bibinfo {year} {2020}{\natexlab{a}})}\BibitemShut
  {NoStop}%
\bibitem [{\citenamefont {Seipt}\ and\ \citenamefont
  {Thomas}(2023)}]{seipt2023kinetic}%
  \BibitemOpen
  \bibfield  {author} {\bibinfo {author} {\bibfnamefont {D.}~\bibnamefont
  {Seipt}}\ and\ \bibinfo {author} {\bibfnamefont {A.~G.}\ \bibnamefont
  {Thomas}},\ }\bibfield  {title} {\bibinfo {title} {Kinetic theory for
  spin-polarized relativistic plasmas},\ }\href@noop {} {\bibfield  {journal}
  {\bibinfo  {journal} {Physics of Plasmas}\ }\textbf {\bibinfo {volume}
  {30}},\ \bibinfo {pages} {093102} (\bibinfo {year} {2023})}\BibitemShut
  {NoStop}%
\bibitem [{\citenamefont {Qian}\ \emph {et~al.}(2023)\citenamefont {Qian} \emph
  {et~al.}}]{qian2023parametric}%
  \BibitemOpen
  \bibfield  {author} {\bibinfo {author} {\bibfnamefont {Q.}~\bibnamefont
  {Qian}} \emph {et~al.},\ }\bibfield  {title} {\bibinfo {title} {Parametric
  study of the polarization dependence of nonlinear breit--wheeler pair
  creation process using two laser pulses},\ }\href@noop {} {\bibfield
  {journal} {\bibinfo  {journal} {Physics of Plasmas}\ }\textbf {\bibinfo
  {volume} {30}},\ \bibinfo {pages} {103107} (\bibinfo {year}
  {2023})}\BibitemShut {NoStop}%
\bibitem [{\citenamefont {Qian}\ \emph {et~al.}(2025)\citenamefont {Qian} \emph
  {et~al.}}]{qian2025fully}%
  \BibitemOpen
  \bibfield  {author} {\bibinfo {author} {\bibfnamefont {Q.}~\bibnamefont
  {Qian}} \emph {et~al.},\ }\bibfield  {title} {\bibinfo {title} {A fully spin
  and polarization resolved strong field qed algorithm for particle-in-cell
  codes},\ }\href@noop {} {\bibfield  {journal} {\bibinfo  {journal} {arXiv
  preprint arXiv:2511.08929}\ } (\bibinfo {year} {2025})}\BibitemShut {NoStop}%
\bibitem [{\citenamefont {Song}\ and\ \citenamefont
  {Tamburini}(2024)}]{song2024polarized}%
  \BibitemOpen
  \bibfield  {author} {\bibinfo {author} {\bibfnamefont {H.-H.}\ \bibnamefont
  {Song}}\ and\ \bibinfo {author} {\bibfnamefont {M.}~\bibnamefont
  {Tamburini}},\ }\bibfield  {title} {\bibinfo {title} {Polarized qed cascades
  over pulsar polar caps},\ }\href@noop {} {\bibfield  {journal} {\bibinfo
  {journal} {Monthly Notices of the Royal Astronomical Society}\ }\textbf
  {\bibinfo {volume} {530}},\ \bibinfo {pages} {2087} (\bibinfo {year}
  {2024})}\BibitemShut {NoStop}%
\bibitem [{\citenamefont {Gong}\ \emph {et~al.}(2025)\citenamefont {Gong},
  \citenamefont {Hatsagortsyan},\ and\ \citenamefont {Keitel}}]{gong2025spin}%
  \BibitemOpen
  \bibfield  {author} {\bibinfo {author} {\bibfnamefont {Z.}~\bibnamefont
  {Gong}}, \bibinfo {author} {\bibfnamefont {K.~Z.}\ \bibnamefont
  {Hatsagortsyan}},\ and\ \bibinfo {author} {\bibfnamefont {C.~H.}\
  \bibnamefont {Keitel}},\ }\bibfield  {title} {\bibinfo {title}
  {Spin-polarized condensed plasmoids in radiation reaction dominated magnetic
  reconnection},\ }\href@noop {} {\bibfield  {journal} {\bibinfo  {journal}
  {Physical Review Letters}\ }\textbf {\bibinfo {volume} {135}},\ \bibinfo
  {pages} {045101} (\bibinfo {year} {2025})}\BibitemShut {NoStop}%
\bibitem [{\citenamefont {Kumar}\ and\ \citenamefont
  {Bo\v{s}njak}(2020)}]{KumarStarvation2020}%
  \BibitemOpen
  \bibfield  {author} {\bibinfo {author} {\bibfnamefont {P.}~\bibnamefont
  {Kumar}}\ and\ \bibinfo {author} {\bibfnamefont {Z.}~\bibnamefont
  {Bo\v{s}njak}},\ }\bibfield  {title} {\bibinfo {title} {Frb coherent emission
  from decay of alfv\'{e}n waves},\ }\href@noop {} {\bibfield  {journal}
  {\bibinfo  {journal} {Monthly Notices of the Royal Astronomical Society}\
  }\textbf {\bibinfo {volume} {494}},\ \bibinfo {pages} {2385} (\bibinfo {year}
  {2020})}\BibitemShut {NoStop}%
\bibitem [{\citenamefont {N{\"a}ttil{\"a}}\ and\ \citenamefont
  {Beloborodov}(2022)}]{nattila2022heating}%
  \BibitemOpen
  \bibfield  {author} {\bibinfo {author} {\bibfnamefont {J.}~\bibnamefont
  {N{\"a}ttil{\"a}}}\ and\ \bibinfo {author} {\bibfnamefont {A.~M.}\
  \bibnamefont {Beloborodov}},\ }\bibfield  {title} {\bibinfo {title} {Heating
  of magnetically dominated plasma by alfv{\'e}n-wave turbulence},\ }\href@noop
  {} {\bibfield  {journal} {\bibinfo  {journal} {Physical Review Letters}\
  }\textbf {\bibinfo {volume} {128}},\ \bibinfo {pages} {075101} (\bibinfo
  {year} {2022})}\BibitemShut {NoStop}%
\bibitem [{\citenamefont {Boldyrev}\ and\ \citenamefont
  {Loureiro}(2025)}]{BoldyrevStarvation2025}%
  \BibitemOpen
  \bibfield  {author} {\bibinfo {author} {\bibfnamefont {S.}~\bibnamefont
  {Boldyrev}}\ and\ \bibinfo {author} {\bibfnamefont {N.~F.}\ \bibnamefont
  {Loureiro}},\ }\bibfield  {title} {\bibinfo {title} {Tearing-mediated
  alfv\'{e}nic turbulence in a relativistic plasma},\ }\href@noop {} {\bibfield
   {journal} {\bibinfo  {journal} {The Astrophysical Journal}\ }\textbf
  {\bibinfo {volume} {979}},\ \bibinfo {pages} {232} (\bibinfo {year}
  {2025})}\BibitemShut {NoStop}%
\bibitem [{\citenamefont {Arber}\ \emph {et~al.}(2015)\citenamefont {Arber}
  \emph {et~al.}}]{arber2015contemporary}%
  \BibitemOpen
  \bibfield  {author} {\bibinfo {author} {\bibfnamefont {T.}~\bibnamefont
  {Arber}} \emph {et~al.},\ }\bibfield  {title} {\bibinfo {title} {Contemporary
  particle-in-cell approach to laser-plasma modelling},\ }\href@noop {}
  {\bibfield  {journal} {\bibinfo  {journal} {Plasma Physics and Controlled
  Fusion}\ }\textbf {\bibinfo {volume} {57}},\ \bibinfo {pages} {113001}
  (\bibinfo {year} {2015})}\BibitemShut {NoStop}%
\bibitem [{SM()}]{SM}%
  \BibitemOpen
  \href@noop {} {}\bibinfo {howpublished} {See the Supplemental Materials for
  the detailed parameter setup for the PIC simulations, the discussion on the
  radiative spin-flip effect, the theoretical model for magnetic energy
  spectrum in the dissipative scenario, and the extra PIC simulations for
  robustness check. The Supplemental Materials include
  Ref.~\cite{arber2015contemporary,comisso2018particle,di2012extremely,thomas1927kinematics,bargmann1959precession,ridgers2014modelling,Chen_2022,gonoskov2015extended,yokoya2003user,gong2021diagnosing,gong2022deciphering,li2020production,montefiori2026intrinsic,schekochihin2009astrophysical,Boldyrev2012,WanIntermittent2012,Zhou2023,nattila2024radiative,Hakobyan2019pair}}\BibitemShut
  {NoStop}%
\bibitem [{\citenamefont {Thomas}(1927)}]{thomas1927kinematics}%
  \BibitemOpen
  \bibfield  {author} {\bibinfo {author} {\bibfnamefont {L.~H.}\ \bibnamefont
  {Thomas}},\ }\bibfield  {title} {\bibinfo {title} {I. the kinematics of an
  electron with an axis},\ }\href@noop {} {\bibfield  {journal} {\bibinfo
  {journal} {The London, Edinburgh, and Dublin Philosophical Magazine and
  Journal of Science}\ }\textbf {\bibinfo {volume} {3}},\ \bibinfo {pages} {1}
  (\bibinfo {year} {1927})}\BibitemShut {NoStop}%
\bibitem [{\citenamefont {Bargmann}\ \emph {et~al.}(1959)\citenamefont
  {Bargmann}, \citenamefont {Michel},\ and\ \citenamefont
  {Telegdi}}]{bargmann1959precession}%
  \BibitemOpen
  \bibfield  {author} {\bibinfo {author} {\bibfnamefont {V.}~\bibnamefont
  {Bargmann}}, \bibinfo {author} {\bibfnamefont {L.}~\bibnamefont {Michel}},\
  and\ \bibinfo {author} {\bibfnamefont {V.}~\bibnamefont {Telegdi}},\
  }\bibfield  {title} {\bibinfo {title} {Precession of the polarization of
  particles moving in a homogeneous electromagnetic field},\ }\href@noop {}
  {\bibfield  {journal} {\bibinfo  {journal} {Physical Review Letters}\
  }\textbf {\bibinfo {volume} {2}},\ \bibinfo {pages} {435} (\bibinfo {year}
  {1959})}\BibitemShut {NoStop}%
\bibitem [{\citenamefont {N{\"a}ttil{\"a}}(2024)}]{nattila2024radiative}%
  \BibitemOpen
  \bibfield  {author} {\bibinfo {author} {\bibfnamefont {J.}~\bibnamefont
  {N{\"a}ttil{\"a}}},\ }\bibfield  {title} {\bibinfo {title} {Radiative plasma
  simulations of black hole accretion flow coronae in the hard and soft
  states},\ }\href@noop {} {\bibfield  {journal} {\bibinfo  {journal} {Nature
  Communications}\ }\textbf {\bibinfo {volume} {15}},\ \bibinfo {pages} {7026}
  (\bibinfo {year} {2024})}\BibitemShut {NoStop}%
\bibitem [{\citenamefont {Boldyrev}\ and\ \citenamefont
  {Perez}(2012)}]{Boldyrev2012}%
  \BibitemOpen
  \bibfield  {author} {\bibinfo {author} {\bibfnamefont {S.}~\bibnamefont
  {Boldyrev}}\ and\ \bibinfo {author} {\bibfnamefont {J.~C.}\ \bibnamefont
  {Perez}},\ }\bibfield  {title} {\bibinfo {title} {Spectrum of
  kinetic-alfv\'{e}n-turbulence},\ }\href@noop {} {\bibfield  {journal}
  {\bibinfo  {journal} {The Astrophysical Journal Letters}\ }\textbf {\bibinfo
  {volume} {758}},\ \bibinfo {pages} {L44} (\bibinfo {year}
  {2012})}\BibitemShut {NoStop}%
\bibitem [{\citenamefont {Wan}\ \emph {et~al.}(2012)\citenamefont {Wan},
  \citenamefont {Matthaeus}, \citenamefont {Karimabadi}, \citenamefont
  {Roytershteyn}, \citenamefont {Shay}, \citenamefont {Wu}, \citenamefont
  {Daughton}, \citenamefont {Loring},\ and\ \citenamefont
  {Chapman}}]{WanIntermittent2012}%
  \BibitemOpen
  \bibfield  {author} {\bibinfo {author} {\bibfnamefont {M.}~\bibnamefont
  {Wan}}, \bibinfo {author} {\bibfnamefont {W.~H.}\ \bibnamefont {Matthaeus}},
  \bibinfo {author} {\bibfnamefont {H.}~\bibnamefont {Karimabadi}}, \bibinfo
  {author} {\bibfnamefont {V.}~\bibnamefont {Roytershteyn}}, \bibinfo {author}
  {\bibfnamefont {M.}~\bibnamefont {Shay}}, \bibinfo {author} {\bibfnamefont
  {P.}~\bibnamefont {Wu}}, \bibinfo {author} {\bibfnamefont {W.}~\bibnamefont
  {Daughton}}, \bibinfo {author} {\bibfnamefont {B.}~\bibnamefont {Loring}},\
  and\ \bibinfo {author} {\bibfnamefont {S.~C.}\ \bibnamefont {Chapman}},\
  }\bibfield  {title} {\bibinfo {title} {Intermittent dissipation at kinetic
  scales in collisionless plasma turbulence},\ }\href@noop {} {\bibfield
  {journal} {\bibinfo  {journal} {Phys. Rev. Lett.}\ }\textbf {\bibinfo
  {volume} {109}},\ \bibinfo {pages} {195001} (\bibinfo {year}
  {2012})}\BibitemShut {NoStop}%
\bibitem [{\citenamefont {Zhou}\ \emph
  {et~al.}(2023{\natexlab{b}})\citenamefont {Zhou}, \citenamefont {Liu},\ and\
  \citenamefont {Loureiro}}]{Zhou2023}%
  \BibitemOpen
  \bibfield  {author} {\bibinfo {author} {\bibfnamefont {M.}~\bibnamefont
  {Zhou}}, \bibinfo {author} {\bibfnamefont {Z.}~\bibnamefont {Liu}},\ and\
  \bibinfo {author} {\bibfnamefont {N.~F.}\ \bibnamefont {Loureiro}},\
  }\bibfield  {title} {\bibinfo {title} {Spectrum of kinetic-alfvén-wave
  turbulence: intermittency or tearing mediation?},\ }\href@noop {} {\bibfield
  {journal} {\bibinfo  {journal} {Monthly Notices of the Royal Astronomical
  Society}\ }\textbf {\bibinfo {volume} {524}},\ \bibinfo {pages} {5468}
  (\bibinfo {year} {2023}{\natexlab{b}})}\BibitemShut {NoStop}%
\bibitem [{\citenamefont {Comisso}\ and\ \citenamefont
  {Sironi}(2019)}]{ComissoAPJ2019}%
  \BibitemOpen
  \bibfield  {author} {\bibinfo {author} {\bibfnamefont {L.}~\bibnamefont
  {Comisso}}\ and\ \bibinfo {author} {\bibfnamefont {L.}~\bibnamefont
  {Sironi}},\ }\bibfield  {title} {\bibinfo {title} {The interplay of
  magnetically dominated turbulence and magnetic reconnection in producing
  nonthermal particles},\ }\href@noop {} {\bibfield  {journal} {\bibinfo
  {journal} {The Astrophysical Journal}\ }\textbf {\bibinfo {volume} {886}},\
  \bibinfo {pages} {122} (\bibinfo {year} {2019})}\BibitemShut {NoStop}%
\bibitem [{\citenamefont {Thompson}\ and\ \citenamefont
  {Duncan}(2001)}]{ThompsonQuake2001}%
  \BibitemOpen
  \bibfield  {author} {\bibinfo {author} {\bibfnamefont {C.}~\bibnamefont
  {Thompson}}\ and\ \bibinfo {author} {\bibfnamefont {R.~C.}\ \bibnamefont
  {Duncan}},\ }\bibfield  {title} {\bibinfo {title} {The giant flare of 1998
  august 27 from sgr 1900+14. ii. radiative mechanism and physical constraints
  on the source},\ }\href@noop {} {\bibfield  {journal} {\bibinfo  {journal}
  {The Astrophysical Journal}\ }\textbf {\bibinfo {volume} {561}},\ \bibinfo
  {pages} {980} (\bibinfo {year} {2001})}\BibitemShut {NoStop}%
\bibitem [{\citenamefont {Zhang}\ \emph {et~al.}(2019)\citenamefont {Zhang}
  \emph {et~al.}}]{zhang2019detailed}%
  \BibitemOpen
  \bibfield  {author} {\bibinfo {author} {\bibfnamefont {S.-N.}\ \bibnamefont
  {Zhang}} \emph {et~al.},\ }\bibfield  {title} {\bibinfo {title} {Detailed
  polarization measurements of the prompt emission of five gamma-ray bursts},\
  }\href@noop {} {\bibfield  {journal} {\bibinfo  {journal} {Nature Astronomy}\
  }\textbf {\bibinfo {volume} {3}},\ \bibinfo {pages} {258} (\bibinfo {year}
  {2019})}\BibitemShut {NoStop}%
\bibitem [{\citenamefont {Kouveliotou}\ \emph {et~al.}(1993)\citenamefont
  {Kouveliotou} \emph {et~al.}}]{kouveliotou1993recurrent}%
  \BibitemOpen
  \bibfield  {author} {\bibinfo {author} {\bibfnamefont {C.}~\bibnamefont
  {Kouveliotou}} \emph {et~al.},\ }\bibfield  {title} {\bibinfo {title}
  {Recurrent burst activity from the soft $\gamma$-ray repeater sgr 1900+ 14},\
  }\href@noop {} {\bibfield  {journal} {\bibinfo  {journal} {Nature}\ }\textbf
  {\bibinfo {volume} {362}},\ \bibinfo {pages} {728} (\bibinfo {year}
  {1993})}\BibitemShut {NoStop}%
\bibitem [{\citenamefont {H.E.S.S.}\ \emph {et~al.}(2023)\citenamefont
  {H.E.S.S.} \emph {et~al.}}]{hess2023discovery}%
  \BibitemOpen
  \bibfield  {author} {\bibinfo {author} {\bibnamefont {H.E.S.S.}} \emph
  {et~al.},\ }\bibfield  {title} {\bibinfo {title} {Discovery of a radiation
  component from the vela pulsar reaching 20 teraelectronvolts},\ }\href@noop
  {} {\bibfield  {journal} {\bibinfo  {journal} {Nature astronomy}\ }\textbf
  {\bibinfo {volume} {7}},\ \bibinfo {pages} {1341} (\bibinfo {year}
  {2023})}\BibitemShut {NoStop}%
\bibitem [{\citenamefont {Hakobyan}\ \emph {et~al.}(2023)\citenamefont
  {Hakobyan}, \citenamefont {Philippov},\ and\ \citenamefont
  {Spitkovsky}}]{hakobyan2023magnetic}%
  \BibitemOpen
  \bibfield  {author} {\bibinfo {author} {\bibfnamefont {H.}~\bibnamefont
  {Hakobyan}}, \bibinfo {author} {\bibfnamefont {A.}~\bibnamefont
  {Philippov}},\ and\ \bibinfo {author} {\bibfnamefont {A.}~\bibnamefont
  {Spitkovsky}},\ }\bibfield  {title} {\bibinfo {title} {Magnetic energy
  dissipation and $\gamma$-ray emission in energetic pulsars},\ }\href@noop {}
  {\bibfield  {journal} {\bibinfo  {journal} {The Astrophysical Journal}\
  }\textbf {\bibinfo {volume} {943}},\ \bibinfo {pages} {105} (\bibinfo {year}
  {2023})}\BibitemShut {NoStop}%
\bibitem [{\citenamefont {Gonoskov}\ \emph {et~al.}(2022)\citenamefont
  {Gonoskov} \emph {et~al.}}]{gonoskov2022charged}%
  \BibitemOpen
  \bibfield  {author} {\bibinfo {author} {\bibfnamefont {A.}~\bibnamefont
  {Gonoskov}} \emph {et~al.},\ }\bibfield  {title} {\bibinfo {title} {Charged
  particle motion and radiation in strong electromagnetic fields},\ }\href@noop
  {} {\bibfield  {journal} {\bibinfo  {journal} {Reviews of Modern Physics}\
  }\textbf {\bibinfo {volume} {94}},\ \bibinfo {pages} {045001} (\bibinfo
  {year} {2022})}\BibitemShut {NoStop}%
\bibitem [{\citenamefont {Chen}\ and\ \citenamefont
  {Fiuza}(2023)}]{chen2023perspectives}%
  \BibitemOpen
  \bibfield  {author} {\bibinfo {author} {\bibfnamefont {H.}~\bibnamefont
  {Chen}}\ and\ \bibinfo {author} {\bibfnamefont {F.}~\bibnamefont {Fiuza}},\
  }\bibfield  {title} {\bibinfo {title} {Perspectives on relativistic
  electron--positron pair plasma experiments of astrophysical relevance using
  high-power lasers},\ }\href@noop {} {\bibfield  {journal} {\bibinfo
  {journal} {Physics of Plasmas}\ }\textbf {\bibinfo {volume} {30}} (\bibinfo
  {year} {2023})}\BibitemShut {NoStop}%
\bibitem [{\citenamefont {Arrowsmith}\ \emph {et~al.}(2024)\citenamefont
  {Arrowsmith} \emph {et~al.}}]{arrowsmith2024laboratory}%
  \BibitemOpen
  \bibfield  {author} {\bibinfo {author} {\bibfnamefont {C.~D.}\ \bibnamefont
  {Arrowsmith}} \emph {et~al.},\ }\bibfield  {title} {\bibinfo {title}
  {Laboratory realization of relativistic pair-plasma beams},\ }\href@noop {}
  {\bibfield  {journal} {\bibinfo  {journal} {Nature Communications}\ }\textbf
  {\bibinfo {volume} {15}},\ \bibinfo {pages} {5029} (\bibinfo {year}
  {2024})}\BibitemShut {NoStop}%
\bibitem [{\citenamefont {Mercuri-Baron}\ \emph {et~al.}(2025)\citenamefont
  {Mercuri-Baron}, \citenamefont {Mironov}, \citenamefont {Riconda},
  \citenamefont {Grassi},\ and\ \citenamefont {Grech}}]{mercuri2025growth}%
  \BibitemOpen
  \bibfield  {author} {\bibinfo {author} {\bibfnamefont {A.}~\bibnamefont
  {Mercuri-Baron}}, \bibinfo {author} {\bibfnamefont {A.}~\bibnamefont
  {Mironov}}, \bibinfo {author} {\bibfnamefont {C.}~\bibnamefont {Riconda}},
  \bibinfo {author} {\bibfnamefont {A.}~\bibnamefont {Grassi}},\ and\ \bibinfo
  {author} {\bibfnamefont {M.}~\bibnamefont {Grech}},\ }\bibfield  {title}
  {\bibinfo {title} {Growth rate of self-sustained qed cascades induced by
  intense lasers},\ }\href@noop {} {\bibfield  {journal} {\bibinfo  {journal}
  {Physical Review X}\ }\textbf {\bibinfo {volume} {15}},\ \bibinfo {pages}
  {011062} (\bibinfo {year} {2025})}\BibitemShut {NoStop}%
\bibitem [{\citenamefont {Pouyez}\ \emph {et~al.}(2025)\citenamefont {Pouyez},
  \citenamefont {Grismayer}, \citenamefont {Grech},\ and\ \citenamefont
  {Riconda}}]{pouyez2025kinetic}%
  \BibitemOpen
  \bibfield  {author} {\bibinfo {author} {\bibfnamefont {M.}~\bibnamefont
  {Pouyez}}, \bibinfo {author} {\bibfnamefont {T.}~\bibnamefont {Grismayer}},
  \bibinfo {author} {\bibfnamefont {M.}~\bibnamefont {Grech}},\ and\ \bibinfo
  {author} {\bibfnamefont {C.}~\bibnamefont {Riconda}},\ }\bibfield  {title}
  {\bibinfo {title} {Kinetic structure of strong-field qed showers in crossed
  electromagnetic fields},\ }\href@noop {} {\bibfield  {journal} {\bibinfo
  {journal} {Physical Review Letters}\ }\textbf {\bibinfo {volume} {134}},\
  \bibinfo {pages} {135001} (\bibinfo {year} {2025})}\BibitemShut {NoStop}%
\bibitem [{\citenamefont {Reichwein}\ \emph {et~al.}(2025)\citenamefont
  {Reichwein} \emph {et~al.}}]{reichwein2025plasma}%
  \BibitemOpen
  \bibfield  {author} {\bibinfo {author} {\bibfnamefont {L.}~\bibnamefont
  {Reichwein}} \emph {et~al.},\ }\bibfield  {title} {\bibinfo {title} {Plasma
  acceleration of polarized particle beams},\ }\href@noop {} {\bibfield
  {journal} {\bibinfo  {journal} {Reports on Progress in Physics}\ }\textbf
  {\bibinfo {volume} {88}},\ \bibinfo {pages} {117001} (\bibinfo {year}
  {2025})}\BibitemShut {NoStop}%
\bibitem [{Dat()}]{Data}%
  \BibitemOpen
  \href@noop {} {}\bibinfo {howpublished}
  {https://cloud.itp.ac.cn/d/5613971ff29d40298ec5/}\BibitemShut {NoStop}%
\bibitem [{\citenamefont {Ridgers}\ \emph {et~al.}(2014)\citenamefont {Ridgers}
  \emph {et~al.}}]{ridgers2014modelling}%
  \BibitemOpen
  \bibfield  {author} {\bibinfo {author} {\bibfnamefont {C.}~\bibnamefont
  {Ridgers}} \emph {et~al.},\ }\bibfield  {title} {\bibinfo {title} {Modelling
  gamma-ray photon emission and pair production in high-intensity laser--matter
  interactions},\ }\href@noop {} {\bibfield  {journal} {\bibinfo  {journal}
  {Journal of Computational Physics}\ }\textbf {\bibinfo {volume} {260}},\
  \bibinfo {pages} {273} (\bibinfo {year} {2014})}\BibitemShut {NoStop}%
\bibitem [{\citenamefont {Chen}\ \emph
  {et~al.}(2022{\natexlab{b}})\citenamefont {Chen} \emph {et~al.}}]{Chen_2022}%
  \BibitemOpen
  \bibfield  {author} {\bibinfo {author} {\bibfnamefont {Y.-Y.}\ \bibnamefont
  {Chen}} \emph {et~al.},\ }\bibfield  {title} {\bibinfo {title} {Electron
  spin- and photon polarization-resolved probabilities of strong-field qed
  processes},\ }\href@noop {} {\bibfield  {journal} {\bibinfo  {journal} {Phys.
  Rev. D}\ }\textbf {\bibinfo {volume} {105}},\ \bibinfo {pages} {116013}
  (\bibinfo {year} {2022}{\natexlab{b}})}\BibitemShut {NoStop}%
\bibitem [{\citenamefont {Gonoskov}\ \emph {et~al.}(2015)\citenamefont
  {Gonoskov} \emph {et~al.}}]{gonoskov2015extended}%
  \BibitemOpen
  \bibfield  {author} {\bibinfo {author} {\bibfnamefont {A.}~\bibnamefont
  {Gonoskov}} \emph {et~al.},\ }\bibfield  {title} {\bibinfo {title} {Extended
  particle-in-cell schemes for physics in ultrastrong laser fields: Review and
  developments},\ }\href@noop {} {\bibfield  {journal} {\bibinfo  {journal}
  {Physical Review E}\ }\textbf {\bibinfo {volume} {92}},\ \bibinfo {pages}
  {023305} (\bibinfo {year} {2015})}\BibitemShut {NoStop}%
\bibitem [{\citenamefont {Yokoya}\ and\ \citenamefont
  {Chen}(2003)}]{yokoya2003user}%
  \BibitemOpen
  \bibfield  {author} {\bibinfo {author} {\bibfnamefont {K.}~\bibnamefont
  {Yokoya}}\ and\ \bibinfo {author} {\bibfnamefont {P.}~\bibnamefont {Chen}},\
  }\href@noop {} {\bibinfo {title} {User’s manual of cain}} (\bibinfo {year}
  {2003})\BibitemShut {NoStop}%
\bibitem [{\citenamefont {Gong}\ \emph {et~al.}(2021)\citenamefont {Gong},
  \citenamefont {Hatsagortsyan},\ and\ \citenamefont
  {Keitel}}]{gong2021diagnosing}%
  \BibitemOpen
  \bibfield  {author} {\bibinfo {author} {\bibfnamefont {Z.}~\bibnamefont
  {Gong}}, \bibinfo {author} {\bibfnamefont {K.~Z.}\ \bibnamefont
  {Hatsagortsyan}},\ and\ \bibinfo {author} {\bibfnamefont {C.~H.}\
  \bibnamefont {Keitel}},\ }\bibfield  {title} {\bibinfo {title} {Retrieving
  transient magnetic fields of ultrarelativistic laser plasma via ejected
  electron polarization},\ }\href@noop {} {\bibfield  {journal} {\bibinfo
  {journal} {Phys. Rev. Lett.}\ }\textbf {\bibinfo {volume} {127}},\ \bibinfo
  {pages} {165002} (\bibinfo {year} {2021})}\BibitemShut {NoStop}%
\bibitem [{\citenamefont {Gong}\ \emph {et~al.}(2022)\citenamefont {Gong},
  \citenamefont {Hatsagortsyan},\ and\ \citenamefont
  {Keitel}}]{gong2022deciphering}%
  \BibitemOpen
  \bibfield  {author} {\bibinfo {author} {\bibfnamefont {Z.}~\bibnamefont
  {Gong}}, \bibinfo {author} {\bibfnamefont {K.~Z.}\ \bibnamefont
  {Hatsagortsyan}},\ and\ \bibinfo {author} {\bibfnamefont {C.~H.}\
  \bibnamefont {Keitel}},\ }\bibfield  {title} {\bibinfo {title} {Deciphering
  in situ electron dynamics of ultrarelativistic plasma via polarization
  pattern of emitted $\gamma$-photons},\ }\href@noop {} {\bibfield  {journal}
  {\bibinfo  {journal} {Physical Review Research}\ }\textbf {\bibinfo {volume}
  {4}},\ \bibinfo {pages} {L022024} (\bibinfo {year} {2022})}\BibitemShut
  {NoStop}%
\bibitem [{\citenamefont {Li}\ \emph {et~al.}(2020{\natexlab{b}})\citenamefont
  {Li} \emph {et~al.}}]{li2020production}%
  \BibitemOpen
  \bibfield  {author} {\bibinfo {author} {\bibfnamefont {Y.-F.}\ \bibnamefont
  {Li}} \emph {et~al.},\ }\bibfield  {title} {\bibinfo {title} {Production of
  highly polarized positron beams via helicity transfer from polarized
  electrons in a strong laser field},\ }\href@noop {} {\bibfield  {journal}
  {\bibinfo  {journal} {Physical Review Letters}\ }\textbf {\bibinfo {volume}
  {125}},\ \bibinfo {pages} {044802} (\bibinfo {year}
  {2020}{\natexlab{b}})}\BibitemShut {NoStop}%
\bibitem [{\citenamefont {Montefiori}\ \emph {et~al.}(2026)\citenamefont
  {Montefiori} \emph {et~al.}}]{montefiori2026intrinsic}%
  \BibitemOpen
  \bibfield  {author} {\bibinfo {author} {\bibfnamefont {S.}~\bibnamefont
  {Montefiori}} \emph {et~al.},\ }\bibfield  {title} {\bibinfo {title}
  {Intrinsic nonlocality of spin-and polarization-resolved probabilities in
  strong-field quantum electrodynamics},\ }\href@noop {} {\bibfield  {journal}
  {\bibinfo  {journal} {arXiv preprint arXiv:2603.11148}\ } (\bibinfo {year}
  {2026})}\BibitemShut {NoStop}%
\bibitem [{\citenamefont {Hakobyan}\ \emph {et~al.}(2019)\citenamefont
  {Hakobyan}, \citenamefont {Philippov},\ and\ \citenamefont
  {Spitkovsky}}]{Hakobyan2019pair}%
  \BibitemOpen
  \bibfield  {author} {\bibinfo {author} {\bibfnamefont {H.}~\bibnamefont
  {Hakobyan}}, \bibinfo {author} {\bibfnamefont {A.}~\bibnamefont
  {Philippov}},\ and\ \bibinfo {author} {\bibfnamefont {A.}~\bibnamefont
  {Spitkovsky}},\ }\bibfield  {title} {\bibinfo {title} {Effects of synchrotron
  cooling and pair production on collisionless relativistic reconnection},\
  }\href@noop {} {\bibfield  {journal} {\bibinfo  {journal} {The Astrophysical
  Journal}\ }\textbf {\bibinfo {volume} {877}},\ \bibinfo {pages} {53}
  (\bibinfo {year} {2019})}\BibitemShut {NoStop}%
\end{thebibliography}%

\black{\section{End Matter}

\textbf{The explanation of the nontrivial power index $p_1>0$:} To unveil the reason of the nontrivial $p_1>0$, we reexamine the magnetic energy evolution in the presence of radiation-induced dissipation. 
Under the quasi-steady state, energy transfer across scales balancing dissipation yields~\cite{SM}
\begin{equation}\label{eq:steady_spectral_eq}
  \frac{d\varepsilon_\lambda}{d\ln\lambda} \sim  -\gamma_{\rm diss} E_\lambda,
\end{equation}
where $\varepsilon_\lambda \sim B_\lambda^3/\lambda$ is the $\lambda$-dependent energy flux~\cite{Boldyrev2012}, $\gamma_{\rm diss}$ is the radiation-induced dissipation rate, and $E_\lambda\sim B_\lambda^2 \lambda$ is the energy of magnetic fluctuation $B_{\lambda}$ at scale $\lambda$. Considering the dissipation rate being in a form of $ \gamma_{\rm diss} \propto  B_\lambda^\beta $, one obtains $ (3\alpha-1) \lambda^{3\alpha-1} \sim -\mathcal{C} \lambda^{\alpha(\beta+2)+1} $ with a power-law assumption $B_\lambda \propto \lambda^\alpha$, where $\mathcal{C}$ is a constant. Balancing the powers of $\lambda$ gives $\alpha =2/(1-\beta)$ and $E_\lambda \sim \lambda^{1-4/(\beta-1)}$. Thus, the magnetic energy spectrum reads
\begin{equation}\label{eq:magnetic_spectrum}
  E(k_r)dk_r \sim k_r^{-2+4/(\beta-1)} dk_r.
\end{equation}
For a scenario with strong synchrotron radiation, $\gamma_{\rm diss} \sim \tau_{\rm cool}^{-1} \sim (d\gamma_e/dt)/\gamma_e \sim  B_\lambda^2$ due to $d\gamma_e/dt \sim 2r_0^2\gamma_e^2 B_\lambda^2/3m_e c$ (i.e. $\beta=2$) with $r_0$ being the classical electron radius, and therefore Eq.\eqref{eq:magnetic_spectrum} reads 
\begin{equation}
  E(k_r)dk_r \sim k_r^{2} dk_r.
\end{equation}
This derivation provides an intuitive explanation for the power index $ p_1\sim1.57$ induced by strong radiative dissipation [Fig.~\ref{fig:fig2_fft_R_spin}(e)].}

\end{document}